\documentclass[aps,prb,twocolumn,showpacs,floatfix,superscriptaddress,floatfix]{revtex4-2}
\usepackage{natbib}
\usepackage{bm}
\usepackage[normalem]{ulem}
\usepackage{lipsum}
\newcommand{\ex}[1]{\langle#1\rangle}

\PassOptionsToPackage{table,dvipsnames}{xcolor}
\usepackage{graphicx}
\usepackage{color}
\usepackage{amsmath}
\usepackage{amssymb}
\usepackage[utf8]{inputenc}
\usepackage{comment}
\usepackage{dcolumn}
\usepackage{bm}
\usepackage{hyperref}
\usepackage{soul}
\usepackage{footmisc}
\usepackage{xfrac}

\usepackage{ulem}
\usepackage{amsfonts}
\usepackage{verbatim}
\usepackage{stackrel}
\usepackage[capitalize]{cleveref}
\usepackage{tikz}
\usetikzlibrary{shapes.geometric, arrows}
\usetikzlibrary{decorations.markings}
\usetikzlibrary{decorations.pathmorphing}
\usetikzlibrary{decorations.pathreplacing}

\graphicspath{{/Images}}
\usepackage{float}
\newcolumntype{P}[1]{>{\centering\arraybackslash}p{#1}}
\definecolor{tdgreen}{rgb}{0,0.4,0}
\definecolor{tlgreen}{rgb}{0,0.7,0}
\definecolor{tpurple}{RGB}{103, 0, 31}
\definecolor{tbrown}{RGB}{127, 39, 4}
\definecolor{tviolet}{RGB}{73, 0, 106}
\definecolor{tgreen}{RGB}{1, 70, 54}
\definecolor{clblue}{rgb}{0.050, 0.35, 0.90}

\newcommand{\Rmnum}[1]{\expandafter\@slowromancap\romannumeral  #1@}

\usepackage{bbold}
\usepackage{color}

\makeatletter
\allowdisplaybreaks
\makeatother

\newcommand{\dd}[2]{\frac{\delta #1}{\delta #2}}
\newcommand{\xc}[0]{\text{xc}}

\begin{document}

\title{Dynamical exchange-correlation potential formalism for spin-$\frac{1}{2}$ Heisenberg\\ and Hubbard 
chains: the antiferromagnetic/half-filled case}

\author{Zhen Zhao}
\email{Zhen.Zhao@teorfys.lu.se}
 \affiliation{Division of Mathematical Physics and ETSF, Lund University, PO Box 118, 221 00 Lund, Sweden}
 \author{Claudio Verdozzi}
 \email{Claudio.Verdozzi@teorfys.lu.se}
 \affiliation{Division of Mathematical Physics and ETSF, Lund University, PO Box 118, 221 00 Lund, Sweden}
 \author{Ferdi Aryasetiawan}
 \email{Ferdi.Aryasetiawan@teorfys.lu.se}
 \affiliation{Division of Mathematical Physics and ETSF, Lund University, PO Box 118, 221 00 Lund, Sweden}

\date{\today}

\begin{abstract}
The exchange-correlation potential formalism previously introduced and applied to the one-dimensional Hubbard model has been extended to spin systems and applied to the case of the one-dimensional antiferromagnetic  spin$-\frac{1}{2}$ Heisenberg model. Within the spin exchange-correlation potential formulation,  
a new sum rule for spin-systems is derived. The 
exchange-correlation potential for the Heisenberg model is extrapolated from exact diagonalization results of small antiferromagnetic Heisenberg clusters. This procedure is also employed to revisit and computationally improve the previous investigation of the exchange-correlation potential of the half-filled Hubbard model, which was based on the exchange-correlation potential of the dimer. Numerical comparisons with exact benchmark calculations for both the Heisenberg and the Hubbard models indicate that, starting from the exchange-correlation potential of a finite cluster, the extrapolation procedure yields a one-particle spectral function with favorable accuracy at a relatively low computational cost. 
In addition, a comparison between the ground state energies for the one-dimensional Hubbard and Heisenberg models displays how  the well known similarity in behavior of the two models at large interactions manifests within the exchange-correlation potential formalism.
\end{abstract}

\maketitle
\section{Introduction}
Lattice models, in spite of their apparent simplicity, can be very valuable to reveal important features in low dimensional and highly correlated quantum systems. This certainly is the case of two highly paradigmatic models of condensed matter physics,
namely the Hubbard \citep{hubbard1963} and spin-$\frac{1}{2}$ quantum Heisenberg models \citep{heisenberg}.

For several decades, these two models have been a testground for new theoretical and computational methods  \citep{giamarchi2004quantum,essler2005,arovas2022}. Notably, they have been used to describe phenomena such as the Mott transition \citep{imada1998}, high $T_\text{c}$ superconductivity\citep{keimer2015}, quantum spin liquids \citep{savary2017}, and quantum entanglement \citep{horodecki2009,scheie2021}. Furthermore, via suitable parameterization from first-principles ground-state calculations, they have also been used to describe the dynamical behavior of real materials, 
which is experimentally measurable via, e.g.,
neutron scattering and angle-resolved photoemission spectroscopy. 
This model approach is very useful when
first-principles descriptions are too complicated to perform.
(see e.g. \citep{motoyama1996,tennant1993,mourigal2013,scheie2022}).

There are a number of approaches of increasing sophistication being continuosly developed to solve the Hubbard and Heisenberg models \citep{ deC1962,yang1966,haldane1980general,faddeev1981, muller1981,essler1991,lake2013multispinon,booth2015}.
Exact analytical solutions remain scarce. In one dimension (1D), both models are integrable and exactly solvable via Bethe ansatz \citep{bethe,shastry1986}. Yet, exact analytic treatments for higher dimensional or even extended 1D systems (e.g., with next-nearest-neighbor coupling) are in general not available. As it happens, already in 1D not all quantities of interest can be accessed: the Bethe ansatz provides information about the energy dispersion \citep{lieb1968,ovchinnikov1970} but not, for example, the spectral weight, one of the more interesting quantities to consider when studying dynamical correlations, which are usually directly connected to experimental results.

On the numerical side, several approaches can be  suitably employed for both models, such as Exact diagonalization (ED) \citep{tohyama1995}, Quantum Monte Carlo (QMC) \citep{suzuki1977monte,lyklema1982quantum,sandvik1992}, and Density Matrix Renormalization Group (DMRG) \citep{white1992density,white1993density,jeckelmann2002}, 
to name a few. ED gives exact and complete information about the system, but is restricted to small systems, thus unable to capture the thermodynamic limit features. DMRG and QMC are applicable to fairly large systems and with high accuracy in 1D \citep{preuss1994,pereira2012,ido2020}, but for higher dimensions the computational cost increases rapidly  \citep{varney2009,yang2016,vaezi2021}.

Density Functional Theory (DFT) \citep{HK1964,KS1965,runge1984,
jones1989,onida2002}, a standard methodology for first-principles treatment of materials, has also been used to study the two models, \citep{CAPELLE2013}, via direct adaptation and application of the lattice case \citep{lopez2004,verdozzi2008,karlsson2011,Carrascal2015,qin2022},
to calculate the model parameters from first-principles (e.g., Hubbard $U$ \citep{dederichs1984,gunnarsson1989,pickett1998} and Heisenberg $J$ \citep{rudra2006}), but also to use model results as input to realistic calculations \citep{anisimov1991}.  Although formally exact, DFT in practice requires approximations for the exchange-correlation 
energy  \citep{cohen2008}. 

The local-density approximation (LDA) and its extension to local-spin-density approximation (LSDA) are widely used in DFT \citep{jones1989,vladimir1997,kohn1999}. L(S)DA  successfully describes many materials, but does not perform well in strongly correlated systems, and  
 much effort has been devoted to improving it. With focus on model lattice systems, one way is to use the exact Bethe ansatz solution of the Hubbard model to approximate the correlation energy of an inhomogeneous lattice system \citep{lima2003}. A similar employment of DFT 
has also been considered for the Heisenberg model \citep{CapelleHeise}. What is noteworthy about these
L(S)DA approaches when applied to the Hubbard  and Heisenberg models is that the exchange-correlation term
has information about the lattice structure and dimensionality of the system.

From a different perspective, a formalism based on the dynamical exchange-correlation potential (Vxc) was recently introduced \citep{aryasetiawan2022time}.
The formalism is not limited by system size, system dimensionality, or type and range of the interaction, and 
it is thus useful to describe electronic and magnetic structures in general situations.
A main feature of the dynamical Vxc formulation is that the coupling between the dynamical Vxc and the Green function occurs as a direct product in space and time. In contrast, the self-energy, which is traditionally used to calculate the Green function, acts on the Green function as a convolution in space and time. 

As a first application of the framework, 
the lattice one-particle Green function of the infinite 1D Hubbard chain was determined \citep{aryasetiawan2022time,tor2022} using an extrapolation scheme, starting from the dynamical Vxc of the Hubbard dimer as input. 
In spite of the simplicity of the approximation used and the low computational load, the scheme provides estimates of the band gap and spectral function in favorable agreement with the results obtained
from the Bethe ansatz and the Dynamical Density Matrix Renormalization Group (DDMRG) \citep{tdmrg2007}. 
One general conclusion from this investigation is that the Vxc formalism provides a simple picture of the one-electron spectrum: for a given momentum, a time-independent term in Vxc together with the kinetic energy term determine the main peak of the spectral function, 
while a time-dependent term in the form of an exponential couples the Green functions with different momenta 
and generates incoherent structures or satellite peaks. 
The energy variable appearing in the 
exponent can be understood as the main bosonic excitations of the system. 

More recently, as a step towards the study of realistic systems, 
the Vxc of the homogeneous electron gas was
calculated within the random-phase approximation \citep{ferdi2023} with the long-term aim of constructing
the Vxc as a universal functional of the ground-state density within
the local-density approximation.

\section{This work, and plan of the paper}

In this work, the Vxc framework is extended to spin systems, more specifically to the 
1D Heisenberg model.
The Vxc-based equation of motion and the sum rule for the spin exchange-correlation hole are derived.  
Furthermore, the extrapolation scheme employed in the previous work for the 1D Hubbard chain is adopted.
The essential idea of the extrapolation scheme is to start from the Vxc
of a finite cluster (kernel), which can be calculated accurately using an exact diagonalization method or
other methods such as the density-matrix renormalization group. By a suitable extrapolation, this is then used to determine the Green function of the corresponding lattice model. The spin Vxc framework within
the extrapolation scheme is applied to calculate the spectral functions of the 1D spin$-\frac{1}{2}$ antiferromagnetic (AFM) Heisenberg model in the thermodynamic limit, starting from the spin Vxc of small clusters.

In addition, the 1D Hubbard chain is revisited.  
In the previous work, the Hubbard dimer
was the kernel, which was used to calculate the Green function of the 1D Hubbard chain.
In this work, in order to improve the quality of the starting Vxc, the cluster size is enlarged so that additional
information arising from interactions beyond nearest-neighbor is captured. The improved Vxc is then used
to calculate the Green function of the half-filled 1D Hubbard chain.

To summarize, the main outcomes of the present work are:
(i) derivation of the Vxc-based equation of motion and the sum rule of the spin exchange-correlation hole for the 1D Heisenberg model, which can be readily generalized to other spin systems; 
(ii) calculations of the spinon Green function for the 1D AFM Heisenberg lattice by extrapolating from a finite-cluster spinon Vxc; 
(iii) improved treatment of the Vxc of the half-filled 1D Hubbard lattice from the previous work by using as kernel a Vxc from a finite cluster; 
(iv) illustration on 
how in the Vxc formalism the well known large-$U$ limit (where results from the Hubbard model match those from the AFM Heisenberg one) is recovered.

The plan of the paper is as follows: in Section \ref{sec:formalism}, we review briefly the general Vxc formalism. Then, in section \ref{sec:heis} and \ref{sec:heis_chain} we extend and apply the approach to the 
1D AFM Heisenberg model. Specifically, in section \ref{sec:heis_4} and \ref{sec:heis_extra}, 
we derive an analytic expression for the spinon Vxc for a four-site chain, and compute the lattice dynamical structure factor by extrapolating the finite cluster Vxc to the infinite case. In section \ref{sec:hub}, we revisit the 1D Hubbard
model and compute the exact Vxc of a finite cluster larger than the dimer, 
with which we improve previous results in the infinite chain limit. In Section \ref{sec:compare} we discuss Vxc from a comparative perspective, addressing
the ground-state energy for both the 1D AFM Heisenberg model and the half-filled 1D Hubbard model in the large $U$ limit. Finally, in Section \ref{sec:con} we provide some conclusive remarks and an outlook.

\section{General formalism and application to the Heisenberg chain}
\label{sec:formalism}

For a system with a one-body term and two-body interactions, the Hamiltonian reads
\begin{eqnarray}
\label{eq:H_gen}
&\hat{H}=\int dr \hat{\psi}^{\dagger}(r)h^0(r)\hat{\psi}(r)\nonumber \\
&+\frac{1}{2}\int dr_1dr_2\hat{\psi}^{\dagger}(r_1)\hat{\psi}^{\dagger}(r_2)v(r_1,r_2)\hat{\psi}(r_2)\hat{\psi}(r_1),\nonumber \\
\end{eqnarray}
where $\hat{\psi}(r)$ is the fermionic field operator and $r=(\mathbf{r},\sigma)$ is a combined space and spin variable. 
The time-ordered Green function is defined in the Heisenberg picture as
\begin{equation}
iG(1,2):=\ex{\mathcal{T}\hat{\psi}(1)\hat{\psi}^{\dagger}(2)},
\end{equation}
where the argument numbers label the space-time $1:=(r_1,t_1)$, $\ex{.}$ denotes the zero-temperature ground-state expectation value, and $\mathcal{T}$ is the time-ordering symbol.
The equation of motion in the Vxc formalism is given by \citep{aryasetiawan2022time}
\begin{equation}
[i\partial_{t_1}-h(r_1)-V^{\text{xc}}(1,2)]G(1,2)=\delta(1-2),
\end{equation}
where the single-particle term 
\begin{equation}
h(r)=h^0(r)+V^{\text{H}}(r)
\end{equation}
contains the Hartree potential 
\begin{eqnarray}
V^{\text{H}}(r)=\int dr'v(r,r')\ex{\hat{\psi}^{\dagger}(r')\hat{\psi}(r')}
\end{eqnarray}
The Vxc reproduces the interaction term containing a special case of the two-particle Green function
, i.e.,
\begin{eqnarray}
\label{eq:Vxc_gen}
V^{\text{xc}}(1,2)iG(1,2)=\int d3v(1,3)\ex{\mathcal{T}\hat{\psi}^\dagger(3)\hat{\psi}(3)\hat{\psi}(1)\hat{\psi}^{\dagger}(2)}\nonumber\\
-V^{\text{H}}(1)iG(1,2),\nonumber\\
\end{eqnarray}
For fermion field operators and in the presence of Coulomb interactions, the bare exchange part of Vxc can be obtained by considering the lowest order of  the first term on the RHS of Eq.~\eqref{eq:Vxc_gen},
\begin{eqnarray}
V^\text{x}(1,2)iG(1,2)=-\int d3v(1-3)G(1,3)G(3,2).
\end{eqnarray}

\subsection{Spin-spin interactions}
\label{sec:heis}
For systems with spin-spin interactions, an observable of central interest is the spin dynamical structure factor,
whose longitudinal and transverse terms are
\begin{eqnarray}
\mkern-30mu S^{zz}(k,\omega)=\frac{1}{N}\sum_{pq}\int dt\ex{\hat{S}_p^z(t)\hat{S}_q^z(0)}e^{i\omega t}e^{-ik(p-q)}
\end{eqnarray}
and
\begin{eqnarray}
\mkern-30mu S^{+-}(k,\omega)=\frac{1}{N}\sum_{pq}\int dt\ex{\hat{S}_p^+(t)\hat{S}_q^-(0)}e^{i\omega t}e^{-ik(p-q)},
\end{eqnarray}
where $\hat{S}_p^{z,+,-}(t)$ are the spin field operators in the Heisenberg picture.

For the Hubbard model, the spin dynamical structure factor can be obtained by solving a two-particle Green function
\begin{eqnarray}
G^{(2)}_{ppqq}(t):=\ex{\mathcal{T}\big[\hat{c}_{p\uparrow}^\dagger(t)\hat{c}_{p\downarrow}(t)\big]\big[\hat{c}_{q\downarrow}^\dagger\hat{c}_{q\uparrow}\big]},
\end{eqnarray}
but the equation of motion of the two-particle Green function contains three-particle Green function, and thus is generally difficult to solve. Simplification
is however recovered for large repulsion, where charge transfer becomes less likely and spin correlations can be obtained by studying the AFM Heisenberg model. It is thus  of fundamental and practical interest to  
discuss the Vxc formalism directly for the Heisenberg model. 

The isotropic 1D Heisenberg Hamiltonian with nearest neighbour (NN) exchange coupling is given by
\begin{eqnarray}
\hat{H}^{\text{Heis}}=-J\sum_p\Big[\frac{1}{2}(\hat{S}_p^+\hat{S}_{p+1}^-+h.c.)+\hat{S}_p^z\hat{S}_{p+1}^z\Big].
\end{eqnarray}
where for convenience we use an even total number of sites before taking the thermodynamic limit.
We define the Green function with spin field operators 
\begin{eqnarray}
\label{eq:GF_heis_def}
iG_{pq}(t)=\theta(t)\ex{\hat{S}_{p}^+(t)\hat{S}_{q}^-(0)}
+\theta(-t)\ex{\hat{S}_{q}^-(0)\hat{S}_{p}^+(t)},\nonumber\\
\end{eqnarray}
in which
the Heisenberg $J$ is the analog of the two-particle interaction in Eq.~\eqref{eq:H_gen}.
From the Heisenberg equation of motion for the spin field operators, the equation of motion of the Green function reads 
\begin{eqnarray}
i\partial_t G_{pq}(t)+iF_{pq}(t)=2\delta_{pq}\delta(t)\ex{\hat{S}^z_p}
\end{eqnarray}
where the interaction term is
\begin{equation}
F_{pq}(t)=-\sum_{l}J_{pl}[\ex{p,l;q}-\ex{l,p;q}].
\end{equation}
Here,
\begin{equation}
\ex{l,p;q}:=\ex{\mathcal{T}\hat{S}_l^z(t^+)\hat{S}_p^+(t)\hat{S}_q^-(0)},
\end{equation}
and $J_{pl}=J(\delta_{l,p+1}+\delta_{l,p-1})$ for the 1D NN exchange coupling.
One can define the spin exchange-correlation potential analogous to the charge case
as follows:
\begin{eqnarray}
\mkern-10mu V_{pp,qq}^\xc(t)iG_{pq}(t):=&&F_{pq}(t)-V_p^\text{H}iG_{pq}(t)-\sum_lV_{pl}^\text{F}iG_{lq}(t),\nonumber\\
\end{eqnarray}
where the last two terms on the right-hand side , $V^\text{H}$ and $V^\text{F}$, are the analog of the Hartree and exchange
potentials, respectively:
\begin{eqnarray}
\label{eq:H}
V_p^\text{H}(t)&:=&-\sum_lJ_{pl}\ex{\hat{S}_l^z},\\
\label{eq:F}
V_{pl}^\text{F}(t)&:=&J_{pl}\ex{\hat{S}_p^z}.
\end{eqnarray}
Consequently, a spin correlator $g_{lpq}(t)$ can be defined such that
\begin{eqnarray}
\ex{l,p;q}=iG_{pq}(t)g_{lpq}(t)\ex{\hat{S}_l^z},
\end{eqnarray}
while the spin exchange-correlation hole $\rho^\xc$ is defined as
\begin{eqnarray}
\rho^\xc_{lpq}(t)iG_{pq}(t)&=&
-\ex{l,p;q}+\ex{\hat{S}_l^z}iG_{pq}(t).
\end{eqnarray}
Denoting the total $z$-component of the spin 
by $S^z=\sum_l\ex{\hat{S}_l^z}$, and observing that
\begin{eqnarray}
\sum_l\ex{l,p;q}=\big[\theta(-t)+S^z\big]iG_{pq}(t),
\end{eqnarray}
we can obtain a sum rule for general spin interactions:
\begin{eqnarray}
\label{eq:sum_rule}
\sum_l\rho_{lpq}^\xc(t)=-\sum_l\big[g_{lpq}(t)-1\big]\ex{\hat{S}_l}=-\theta(-t).
\end{eqnarray}
The detailed derivation is provided in Appendix \ref{app:sum_rule}.

In this paper, we consider only the case of AFM coupling, i.e.  $J<0$ so that $S^z=0$. For a translationally invariant system, the Hartree and Fock terms (Eq.~\eqref{eq:H},~\eqref{eq:F}) vanish, and thus the two-spinon Vxc is then
\begin{eqnarray}
V_{pp,qq}^\xc(t)iG_{pq}(t)=-J\sum_{\delta=\pm1}\Big[\ex{p,p+\delta;q}-\ex{p+\delta,p;q}
\Big],\nonumber
\end{eqnarray}
with the corresponding exchange term given by
\begin{eqnarray}
\mkern-30mu F^\text{x}_{pq}(t)&&:=V_{pp,qq}^\text{x}(t)iG_{pq}(t)=\nonumber\\
&&J\Big[G_{p+1,p}(0^-)G_{pq}(t)+G_{p-1,p}(0^-)G_{pq}(t)-\nonumber\\
&&G_{p,p+1}(0^-)G_{p+1,q}(t)-G_{p,p-1}(0^+)G_{p-1,q}(t)\Big].
\end{eqnarray}
%
\subsection{The infinite chain} 
\label{sec:heis_chain}
We specialize now the description to the case of the homogeneous infinite Heisenberg chain, 
where $\ex{S_p^z}\equiv s$ is site-independent due to translational symmetry.
It is convenient to move to the momentum domain, with the Green function and $V^\text{xc}$ defined 
via the Fourier transform as
\begin{eqnarray}
G(k,t)&=&\frac{1}{N}\sum_{pq}G_{pq}(t)e^{-ik(p-q)}\\
V^\xc(k,t)&=&\frac{1}{N^2}\sum_{pq}V^\xc_{pp,qq}(t)e^{-ik(p-q)},
\end{eqnarray}
and where the equation of motion for the Green function becomes
\begin{eqnarray}
\label{eq:eom_heis_k}
i\partial_k G(k,t)-\sum_{k'}V^\xc(k-k',t)G(k',t)=2s\delta(t).
\end{eqnarray}
In the momentum representation, the exchange term becomes
\begin{eqnarray}
F^\text{x}(k,t)&&=\frac{4J}{N}G(k,t)\sin\frac{k}{2}\sum_{k'}G(k',0^-)\sin(k'-\frac{k}{2})\nonumber\\
&& \approx  iJ\lambda|\sin k|G(k,t), \label{eq:Fx}
\end{eqnarray}
where we have neglected the $k'$-dependence in the weight represented by $G(k',0^-)$, performed the sum over $k'$ and
subsumed all the constants into $\lambda$.
To proceed further, the dynamical part of Vxc is separated from the static exchange term $V^\text{s}(k)=F^\text{x}(k,t)/iG(k,t)$, i.e.
\begin{eqnarray}
\sum_{k'}V^\xc(k-k',t)G(k',t)=V^\text{s}(k)G(k,t)\nonumber\\
+Z^\text{sp}(k,t)G(k,t), 
\end{eqnarray}
to finally arrive at the solution to the equation of motion Eq.\eqref{eq:eom_heis_k}:
\begin{eqnarray}
\label{eq:G_sp}
G(k,t)=G(k,0^+)e^{-iV^\text{s}t}e^{-i\int_0^tZ^{\text{sp}}(k,t')dt}.
\end{eqnarray}
In this expression, the ($k$-dependent) static exchange term $V^\text{s}$ determines the main peak of the spectral function, and the dynamical correlation term $Z^\text{sp}(k,t)$ produces the satellite structure. To attain an explicit solution,
it is expedient  to solve for a reference Green function by keeping only the static $V^\text{s}$ term in the equation of motion. This simplified
solution contains the lower boundary of the two-spinon energy dispersion \citep{deC1962} 
\begin{eqnarray}
G^{\text{lb}}(k,\omega)=\frac{1}{\omega-(-J)\lambda|\sin k|}, 
\end{eqnarray}
and permits to determine the constant $\lambda$ in Eq.~\eqref{eq:Fx} from the analytic form of the two-spinon spectrum. 
\subsection{A four-site spin chain}
\label{sec:heis_4}
It is useful to start our discussion of Vxc in finite spin-clusters by considering a four-site chain. This is the minimal cluster (with even number of sites)
in which Vxc is nonzero. Furthermore, it is easy to obtain a compact analytical solution, that
illustrates qualitatively several features present also in larger clusters (in which our solution is numerical in character). 
To illustrate the features of the four-site Vxc, we choose one of its diagonal elements as a representative case, namely
\begin{widetext}
\begin{eqnarray}\label{wideq}
V^{\xc}_{11,11}(t>0)=-J\frac{(\frac{(xy+x)(xy+x+2y)}{a_+^2})f_1+(x^2+x)f_2+(\frac{(xy-3x)(xy-3x+2y-4)}{a_-^2})f_3}{(\frac{xy+x+2y}{a_+})^2f_1+x^2f_2+(\frac{xy-3x+2y-4}{a_-})^2f_3}
\end{eqnarray}
\end{widetext}
where
$ x=1+\sqrt{3}, y=1+\sqrt{2}, a_\pm=\sqrt{8\pm4\sqrt{2}}$, 
and $f_{i=1,2,3}$ are time oscillation factors  determined by the difference between the spin excitation energies and the ground state energy. 
The full details and the explicit forms are given in  Appendix \ref{app:spVxc}, together with other elements of Vxc. 
It is useful at this point to move from site orbitals $\lbrace\varphi_a\rbrace$ to bonding-like ones $\lbrace\phi_\mu\rbrace$.
In analogy to what is done with a Bloch basis in a Hubbard lattice, we set $\phi_\mu = U_{\mu a}\varphi_a,  \varphi_a = U_{a\mu}\phi_\mu$, in which $\mu=A,B,C,D$,  $a=1,2,3,4$ and the $ U$ matrix is
\begin{eqnarray}
U=\frac{1}{2}\left(\begin{array}{cccc}
1 & \;\;\;1 & \;\;\;1 & \;\;\;1\\
1 & -1 & \;\;\;1 & -1\\
1 &  \;\;\;1 & -1 & -1\\
1 & -1 & -1 & \;\;\;1
\end{array}
\right).
\end{eqnarray}
For the Green functions, the transformation reads $G_{\mu\nu}=\sum_{ab}U_{\mu a}G_{ab}U_{b\nu }^*$ and 
$G_{ab}=\sum_{\mu\nu}U_{a\mu}^*G_{\mu\nu}U_{\nu b}$.
One can define
\begin{eqnarray}
V^{\xc}_{\mu\alpha,\beta\nu}(t):=\sum_{mn}U_{\mu m}U_{m\alpha}^*V^{\xc}_{mm,nn}(t)U_{\beta n}U_{n \nu}^*
\end{eqnarray}
such that the equation of motion is now
\begin{eqnarray}
i\partial_tG_{\mu\nu}(t)-\sum_{\alpha\beta}V^{\xc}_{\mu\alpha,\beta\nu}(t)G_{\alpha\beta}(t)=s_{\mu\nu}\delta(t),
\end{eqnarray}
where $ s_{\mu\nu}=2\sum_{pq}U_{\mu p}\ex{S_p^z}\delta_{pq}U_{q\nu}^*$.
Comparing the equation of motion for the diagonal terms $G_{\mu\mu}$,
\begin{eqnarray}
\label{eq:eom_heis_mu}
[i\partial_t-V^{\xc}_{\mu\mu,\mu\mu}]G_{\mu\mu}(t)-\sum_{\gamma\neq\mu}V^{\xc}_{\mu\gamma,\gamma\mu}G_{\gamma\gamma}\nonumber\\
-\sum_{\substack{\gamma\neq\delta}}V^{\xc}_{\mu\gamma,\delta\mu}(t)G_{\gamma\delta}(t)=s_{\mu\mu}\delta(t)
\end{eqnarray}
to the infinite-chain equation of motion Eq.~\eqref{eq:eom_heis_k}, we note the following: i) $G_{\mu\mu}$ maps to $G(k)$; ii) the contribution from fully off-diagonal terms $V^\xc_{\mu\nu,\delta\mu}$ should be negligible; iii) $V^{xc}_{\mu\gamma,\gamma\mu}$, which maps to $V(k)$, depends only on the difference of $\mu,\gamma$; iv) the weights of the higher excitation term $f_3$ are relatively small. 

\begin{figure}
\centering
\includegraphics[scale=0.4]{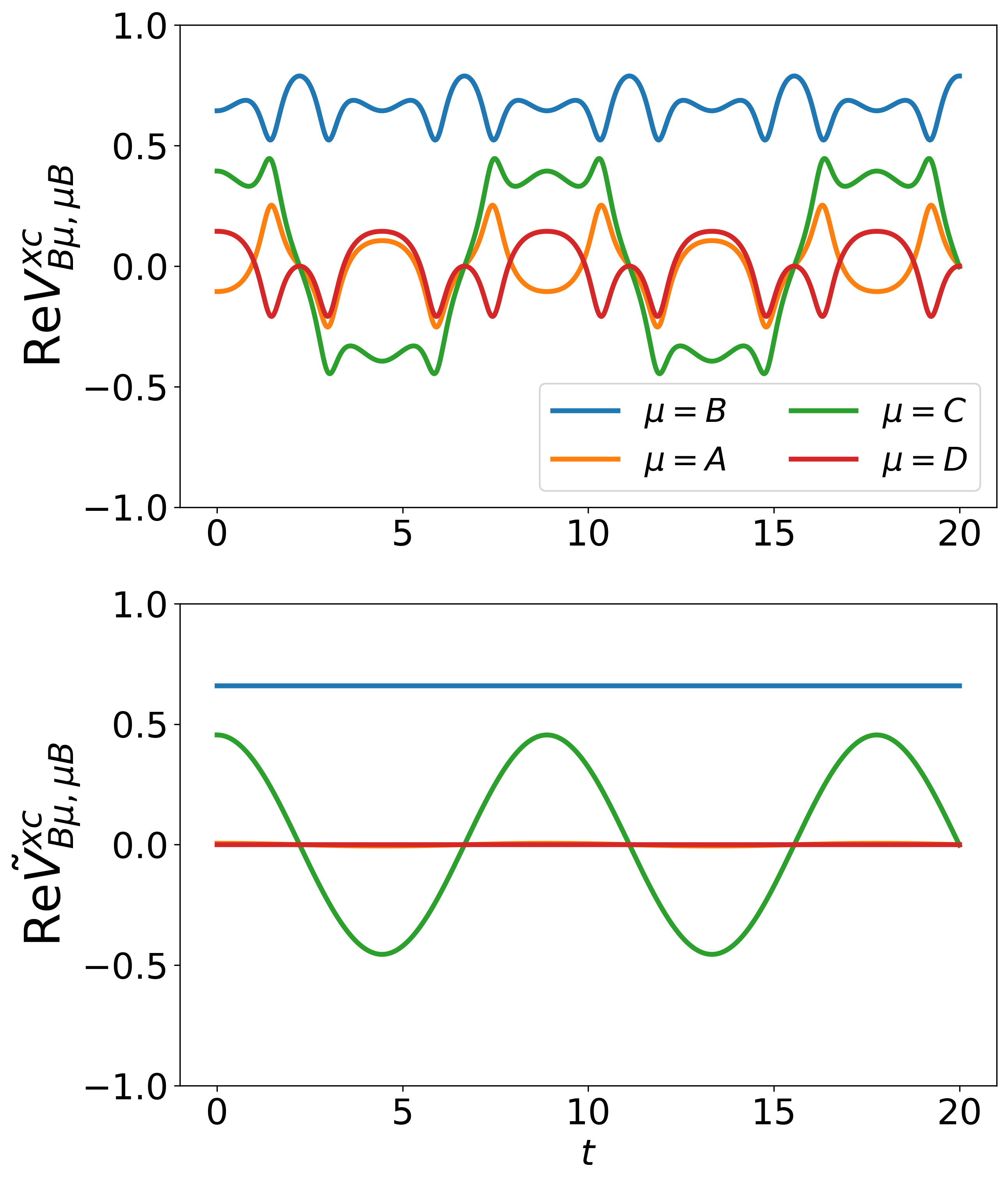}
\caption{Real part of Vxc of four-site spin-$\frac{1}{2}$ AFM Heisenberg chain, in the unit of $|J|$. Top panel: exact. Bottom panel: results when the high-excitation contribution is ignored (see Eqs.~\eqref{eq:Vxc_heis_apprx}-\eqref{beta_eq} and related discussion).}
\label{fig:Vxc_analy}
\end{figure}

According to i-iv) and ignoring the high energy-excitation  contributions from $f_3$, one thus arrives
at an approximate expression for the matrix elements of 
$V^{xc}_{\mu\gamma,\gamma\mu}$:
\begin{eqnarray}
\label{eq:Vxc_heis_apprx}
V_{BB,BB}^{\xc}(t>0)&\approx &-J\alpha\nonumber \\
V_{BC,CB}^{\xc}(t>0)&\approx &-J\beta \exp[\frac{iJt}{\sqrt{2}}],
\end{eqnarray}
whereas $V_{BD,DB}^{\xc}(t>0) \approx 0, V_{BA,AB}^{\xc}(t>0) \approx 0$, in which
\begin{eqnarray}
&\alpha :=\frac{xy+x}{xy+x+2y}=\frac{2x+2}{xy+x+2},\\
&\beta :=\frac{1}{4}(\frac{a_+}{xy+x+2y}+\frac{a_+}{xy+x+2})^2(x^2+x-\alpha x^2), \label{beta_eq}
\end{eqnarray}
and as in \eqref{wideq}, $a_+=\sqrt{8+4\sqrt{2}}$.
The analytic spinon Vxc in the bonding basis and its main excitation approximation are shown in Fig. \ref{fig:Vxc_analy}. Ignoring the high-excitation factor $f_3$ reduces the fine-structure details in Vxc. Consequently, $V^\xc_{BB,BB}$ simplifies to a constant whereas $V^\xc_{BC,CB}$ oscillates with a single frequency and a constant magnitude, and all other components are negligible. 
\vspace{0.5cm}
\subsection{Infinite chain from cluster extrapolation}
\label{sec:heis_extra}
\begin{figure}
\centering
\includegraphics[scale=0.4]{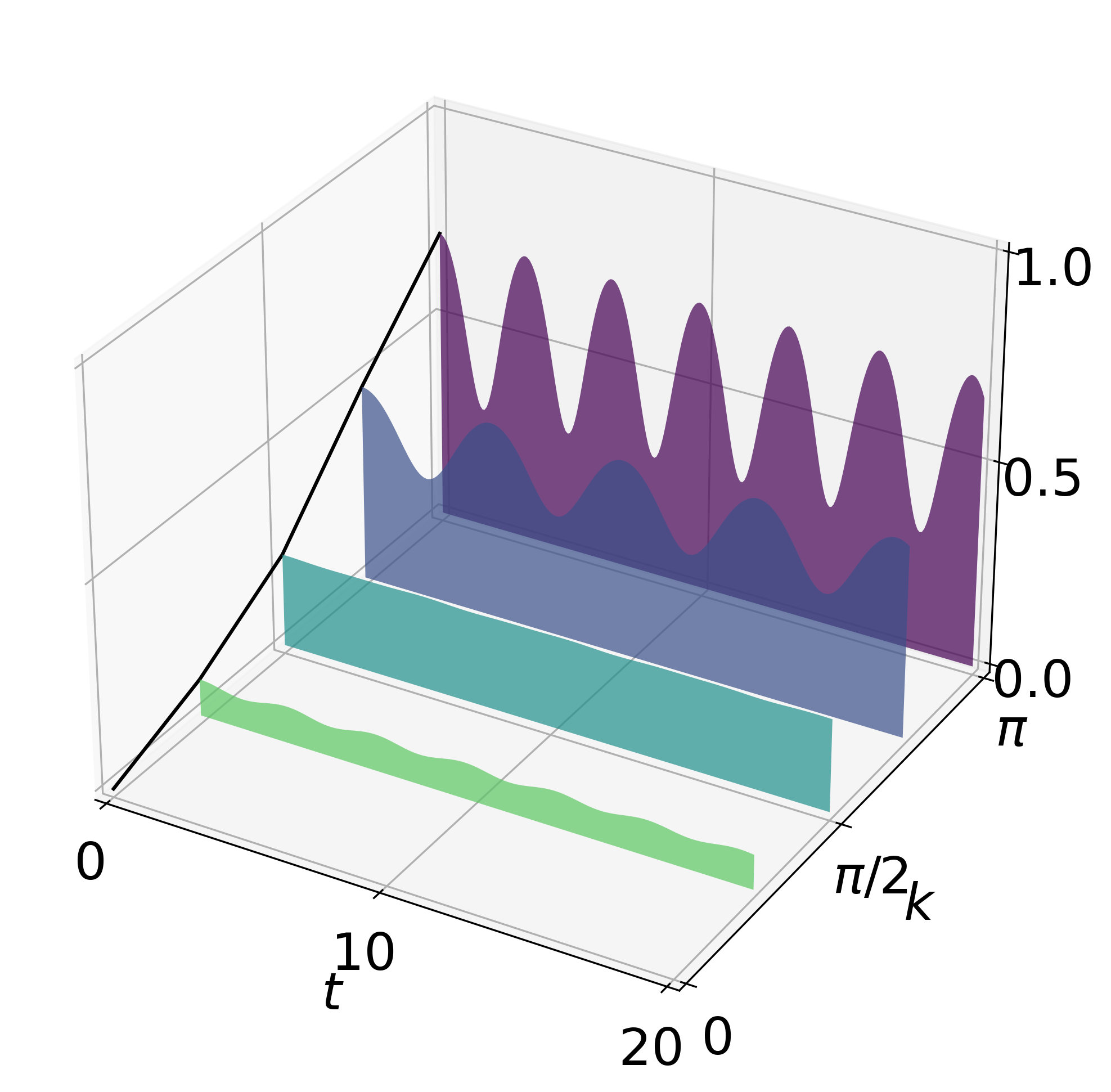}
\includegraphics[scale=0.4]{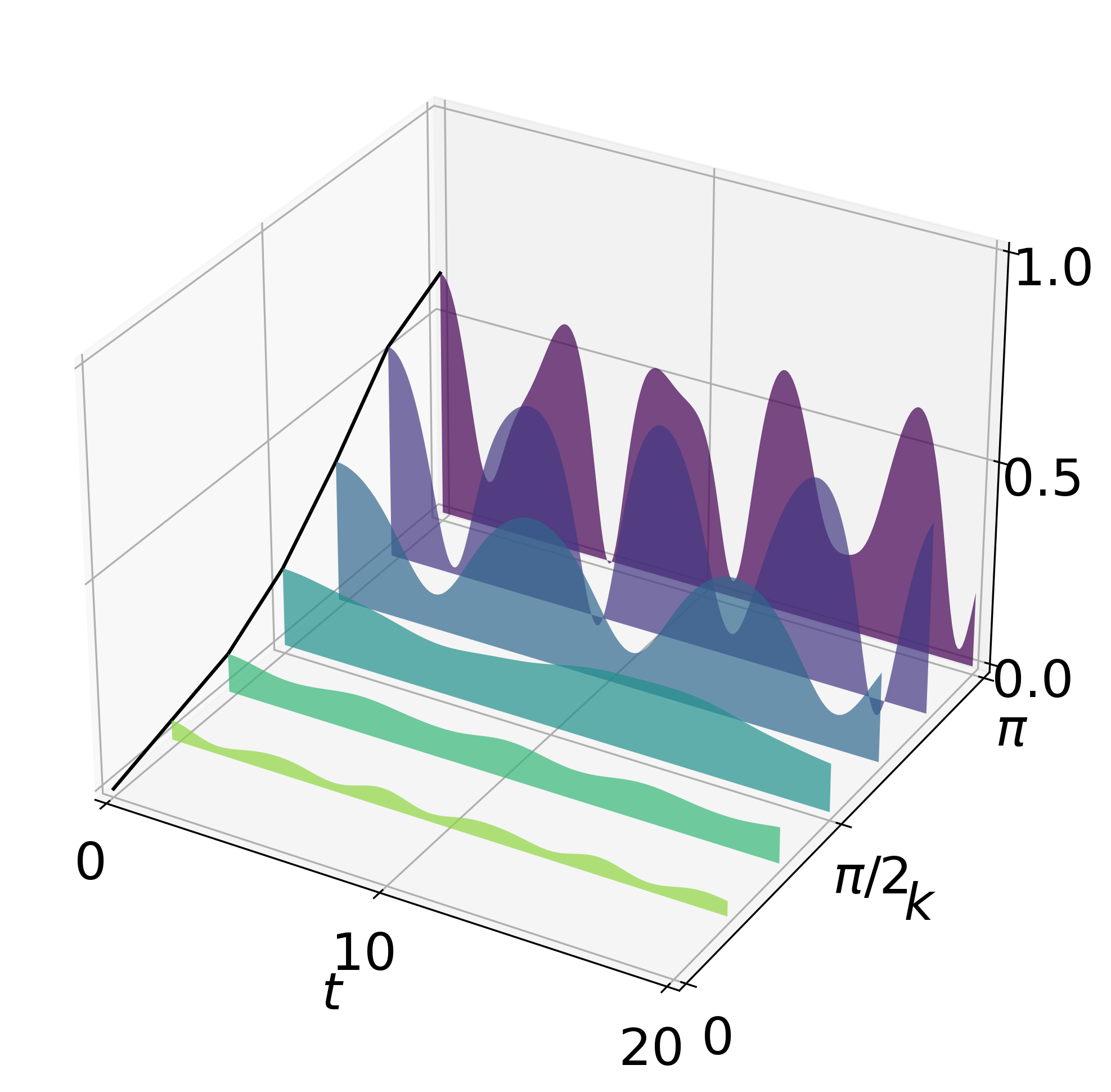}
\caption{Real part of $Z^\text{sp}$ from a  spin-$\frac{1}{2}$ AFM Heisenberg ring. Top (bottom) panel: results for a ring with 8 (12) sites.}
\label{fig:Z_hei}
\end{figure}

In Fig. \ref{fig:Z_hei}, we show $\text{Re}Z^\text{sp}(k,t)$ as obtained from the cluster Vxc discussed in the previous section.

It can be seen that $\text{Re}Z^\text{sp}(k,t)$ oscillates in time around a momentum-dependent term, a behavior that can be understood as 
due to a single quasiparticle-like main excitation. We therefore propose the following ansatz for $Z^\text{sp}$ in the infinite-chain case:
\begin{eqnarray}
\label{eq:sp_quasip}
Z^\text{sp}(k,t)=\mathcal{A}(k)e^{-i\omega^\text{sp}(k) t}+\mathcal{B}(k),
\end{eqnarray}
where the amplitude $\mathcal{A}$, the spinon excitation energy $\omega^\text{sp}$, and the shift term $\mathcal{B}$ all increase as $k$ increases from $0$ to $\pi$. The Green function is given by inserting the ansatz into Eq.\eqref{eq:G_sp}:
\begin{eqnarray}
\label{eq:G_sp_Z}
&&G^\text{sp}(k,t>0)=\nonumber \\&&G^\text{sp}(k,0^+)e^{-i[V^\text{s}(k)+\mathcal{B}(k)]t}
e^{\frac{\mathcal{A}(k)}{\omega^\text{sp}(k)}(e^{-i\omega^\text{sp}(k)t}-1)},
\end{eqnarray}
where the static potential is
$V^\text{s}(k)=-J\pi|\sin k|/2$.

Expanding the last term on the RHS of Eq.\eqref{eq:G_sp_Z} to first order in $e^{-i\omega^\text{sp}(k)t}$, one gets an approximate Green function
\begin{eqnarray}
G^\text{sp}_{(1)}(k,t>0)=G^\text{sp}(k,0^+)e^{-i[V^\text{s}(k)+\mathcal{B}(k)]t}\times\nonumber\\
\big[1+\frac{\mathcal{A}(k)}{\omega^\text{sp}(k)}(e^{-i\omega^\text{sp}(k)t}-1)\big],
\end{eqnarray}
which in the frequency domain becomes
\begin{widetext}
\begin{eqnarray}
G^\text{sp}_{(1)}(k,\omega)=G^\text{sp}(k,0^+)\Big[\frac{1-\frac{\mathcal{A}(k)}{\omega^\text{sp}(k)}}{\omega-[V^\text{s}(k)+\mathcal{B}(k)]}+\frac{\frac{\mathcal{A}(k)}{\omega^\text{sp}(k)}}{\omega-[V^\text{s}(k)+\mathcal{B}(k)+\omega^\text{sp}(k)]}\Big].\label{FinalVxc}
\end{eqnarray}
\end{widetext}
From Eq.~\eqref{FinalVxc}, it can be seen that the main peak position of the dynamical structure factor is given by $V^\text{s}+\mathcal{B}$. The spinon excitation energy $\omega^\text{sp}$ transfers weight from the main peak to higher-energy region resulting in satellite peaks at $V^\text{s}+\mathcal{B}+\omega^\text{sp}$. The relative weight between the main peak and the satellite is determined by the amplitude term $\mathcal{A}$ and the spinon energy $\omega^\text{sp}$. Specifically at $k=\pi$, the finite cluster solution gives nonzero $\mathcal{B}$, which opens a spin gap that does not exist for the spin$-\frac{1}{2}$ lattice. We attribute this to finite-size effects, and thus we adjust $\mathcal{B}$ to a smaller value in our extrapolation.

\begin{figure}
\centering
\includegraphics[scale=0.35]{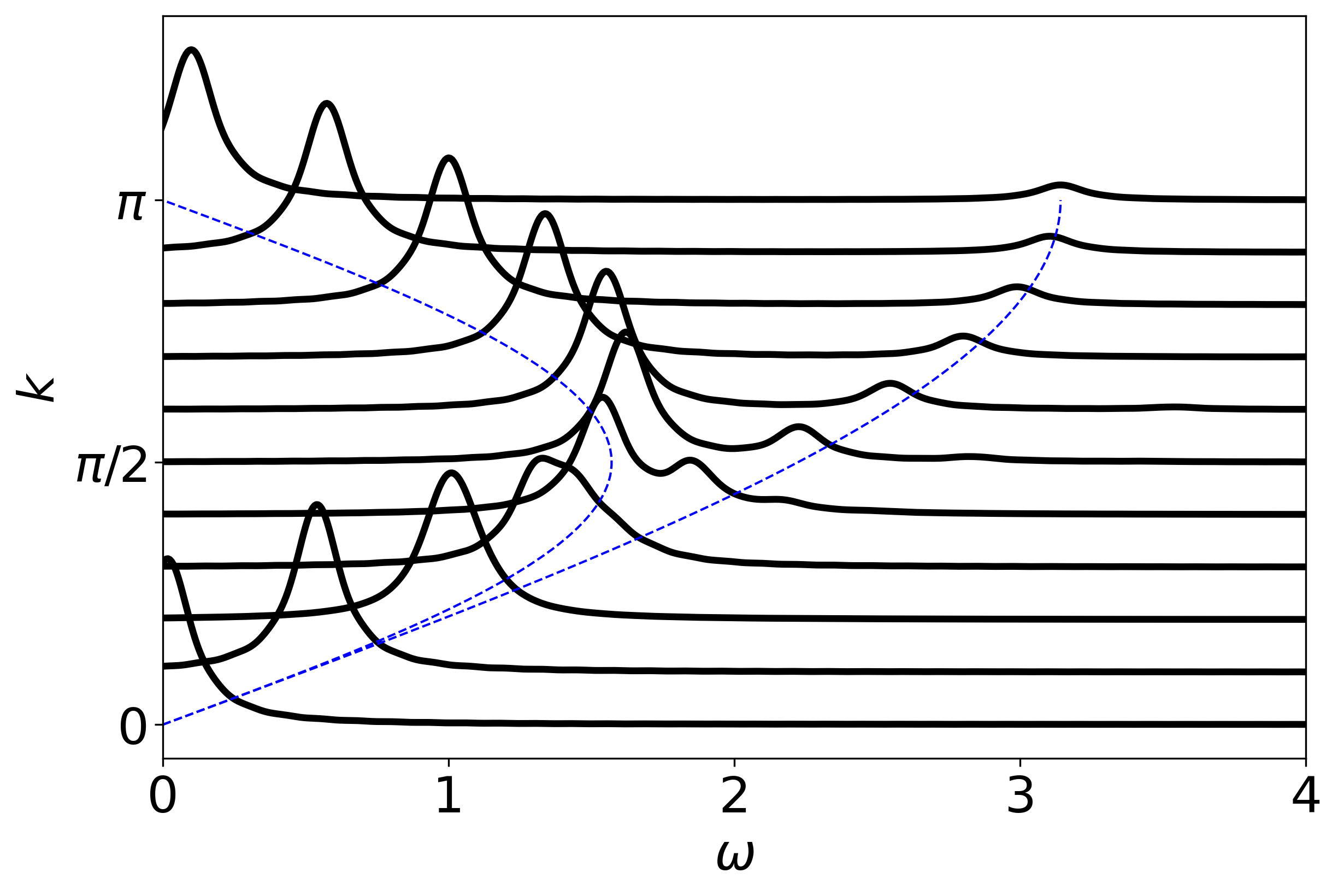}
\includegraphics[scale=0.35]{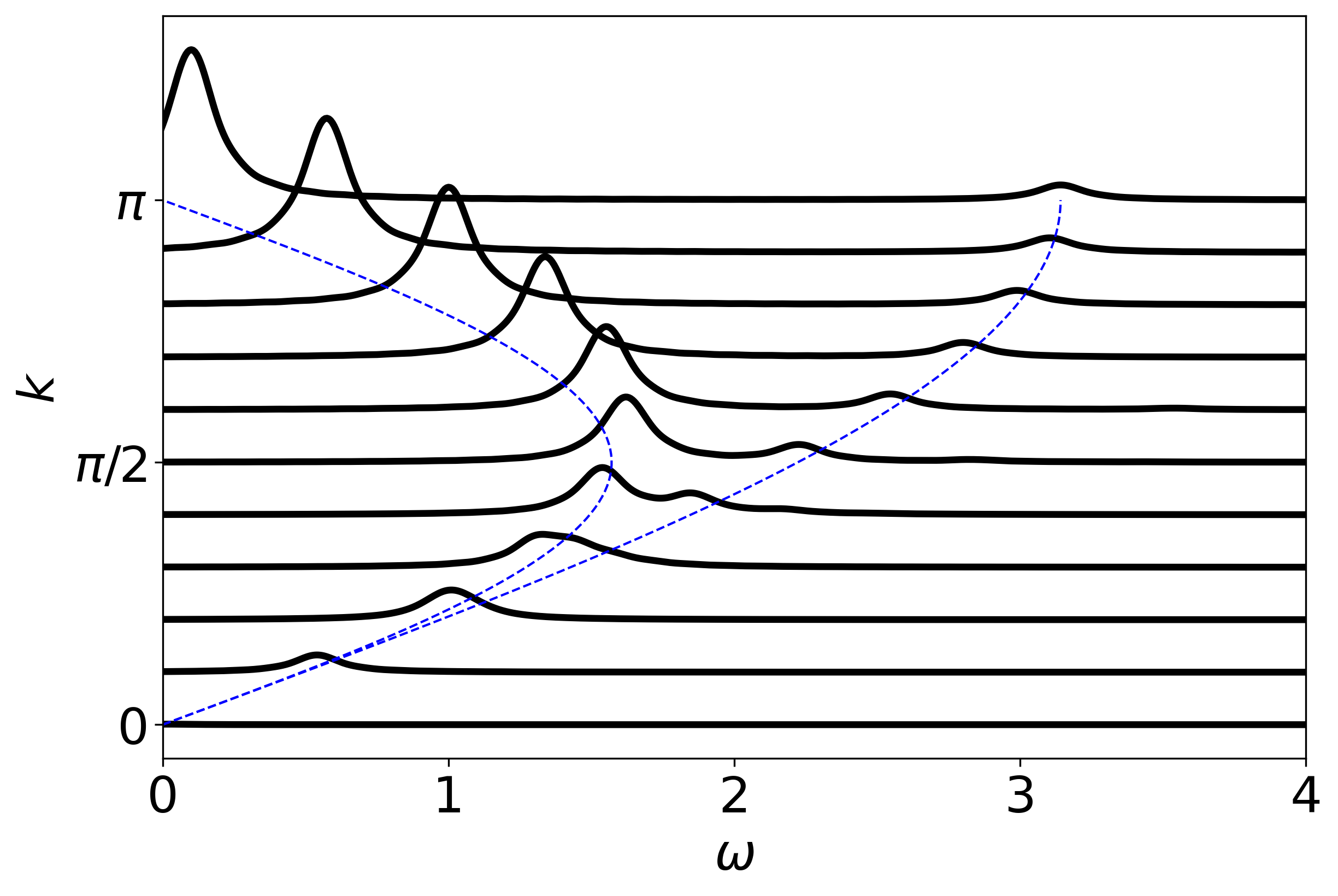}
\caption{Dynamic structure factor of 1D spin-$\frac{1}{2}$ AFM Heisenberg lattice calculated with Vxc, with broadening 0.1. Top panel: weight factor $G(k,t=0)$ considered as unit. Bottom panel: weight renormalised with cluster $G(k,0^+)$. The blue dashed curves are the boundaries for two-spinon processes.}
\label{fig:Sqw_hei}
\end{figure}

Based on our discussion so far, we now present the lattice case obtained by extrapolating the cluster Vxc. With $Z^\text{sp}$ obtained via ED
from a twelve-site cluster, we estimate the parameters $\mathcal{A},\mathcal{B}$ and $G(k,0^+)$ by linear interpolation. The spinon excitation energy is estimated by fitting the cluster $\omega^\text{sp}$ to the two-spinon spectrum boundary,
\begin{eqnarray}
\omega^{\text{sp}}\rightarrow(-J)\pi\big[\sin\frac{k}{2}-\frac{1}{2}|\sin k|\big].
\end{eqnarray}
The longitudinal and transverse spin dynamical structure factors are then calculated from the spinon Green function. Since for a spin isotropic system $S^{zz}$ and $S^{+-}$ 
differ by a constant factor, we only calculate the spectral function of the Green function (Eq.~\eqref{eq:G_sp_Z}), as shown in Fig.~\ref{fig:Sqw_hei} 
 (interestingly, approximating the term $\exp\Big\lbrace\frac{\mathcal{A}}{\omega^\text{sp}}\big[\exp(-i\omega^\text{sp}t)-1\big]\Big\rbrace$ by $1+\frac{\mathcal{A}}{\omega^\text{sp}}\big[\exp(-i\omega^\text{sp}t)-1\big]$ gives no marked changes of the properties of $G^\text{sp}(k,\omega)$).
A notable aspect in the behavior on the spin dynamical structure factor is that both the peak locations and the relative weights are close to the inelastic neutron scattering data from 1D compound KCuF$_3$ \citep{lake2013multispinon}. Coming to more specific features, $S(k,\omega)$ is very small (i.e., close to zero) at small $k$, while, for a generic $k$, most of its spectral weight is concentrated around the main peak and the satellite peak. As $k\rightarrow\pi$, the relative weight  between the main peak and the satellite peak increases and the spectrum with broadening factor 0.1 is gapless.

While providing a good description of the dynamical structure factor for the 1D AFM Heisenberg model, the present implementation of the spin Vxc approach is also subject to 
some limitations. This can be seen by, e.g., comparing the dynamical structure factor from the Vxc approach with the two-spinon lower and upper boundaries (dashed line in Fig.~\ref{fig:Sqw_hei}). It is apparent that the main peak frequency $\omega=V^\text{s}(k)+\mathcal{B}$ is still slightly overestimated. 
To reduce the finite size effects due to a parameter $\mathcal{B}(\pi)$ originating from a twelve-site cluster,  
we set $\mathcal{B}(\pi)$ to be the same as the broadening factor, i.e. about 0.2 (see Fig.~\ref{fig:Z_hei}). However, the actual  
Bethe ansatz value of $\mathcal{B}(\pi)$ should be zero. The overall point is that, to obtain a more accurate dynamical structure factor, and to avoid the finite size effects inherent in the extrapolation from a small cluster, more powerful external methods need to be employed (e.g., the algebraic Bethe ansatz). 

These considerations might reveal weaknesses of the extrapolation procedure. However, it must also be clearly stressed that this implementation of the Vxc approach captures most of the qualitative features of the 1D AFM Heisenberg model with a very low computational load and this central attractive feature of the method is expected to also apply in more challenging situations, e.g.  in higher dimensions, where rigorous references like the Bethe ansatz are not available.

\section{Improving the treatment of the 1D-Hubbard lattice}
\label{sec:hub}
Encouraged by the 1D AFM Heisenberg chain results obtained with a Vxc extrapolated from clusters, 
we now revisit the case of the 1D Hubbard Hamiltonian,
\begin{eqnarray}
\label{eq:def_hub}
\hat{H}^{\text{Hub}}=-\Delta\sum_{p\sigma}[\hat{c}_{p,\sigma}^\dagger\hat{c}_{p+1,\sigma}+h.c.]+U\sum_p\hat{n}_{p\uparrow}\hat{n}_{p\downarrow},
\end{eqnarray}
using also in this case a Vxc obtained from small (Hubbard) clusters.
In Eq.~\eqref{eq:def_hub}, $p=1,2,\cdots,N$ are the site labels (with $N\rightarrow \infty$ eventually), $\sigma=\uparrow,\downarrow$ is the spin label, $\Delta$ is the hopping energy and $U>0$ is the local repulsion. In the site basis, the spin-up channel Green function is
\begin{eqnarray}
\label{eq:GF_hub_def}
\!\!\!\!\!\!\!\!\!\!\!\!G_{pq}(t)=-i\theta(t)\ex{\hat{c}_{p\uparrow}(t)\hat{c}_{q\uparrow}^\dagger(0)}
+i\theta(-t)\ex{\hat{c}_{q\uparrow}^\dagger(0)\hat{c}_{p\uparrow}(t)}
\end{eqnarray}
and the Vxc reads
\begin{eqnarray}
\label{eq:G2_hub}
V_{pp,qq}^\xc(t)iG_{pq}(t)=U\ex{\mathcal{T}\hat{c}_{p\downarrow}^\dagger(t)\hat{c}_{p\downarrow}(t)\hat{c}_{p\uparrow}(t)\hat{c}_{q\uparrow}^\dagger(0)}\nonumber\\
-U\rho_{p\downarrow} iG_{pq}(t),
\end{eqnarray}
where $\rho_{p\downarrow}$ is the spin-down particle density at site $p$. 
The exchange part of Vxc fulfils
\begin{eqnarray}
V_{pp,qq}^\text{x}(t) iG_{pq}(t)=-UG_{pp}(0^-)G_{pq}(t),
\end{eqnarray}
where $G_{pp}(0^-)=i\ex{\hat{c}_{p\uparrow}^\dagger\hat{c}_{p\uparrow}}=i\rho_{p\uparrow}$; thus, the exchange part of Vxc of the Hubbard model is static and cancels the Hartree potential at half-filling, in contrast to the Heisenberg model in which the exchange part depends on the momentum. In general, the exchange part is time-dependent \citep{ferdi2023}.

Written in the momentum domain, the equation of motion for the Hubbard lattice takes the form 
\begin{eqnarray}
\label{eq:eom_hub_k}
[i\partial_k-\varepsilon_k]G(k,t)-\sum_{k'}V^\xc(k-k',t)G(k',t)=\delta(t),\nonumber\\
\end{eqnarray}
where $\varepsilon_k=-2\Delta \cos k$ is the kinetic energy.
Eq.~\eqref{eq:eom_hub_k} shows that the interaction term, expressed as the direct product of Vxc and Green function in space-time domain, is a convolution in the momentum domain. It has been shown \citep{tor2022} that the main peak position of the electron (hole) spectral functions can be described with $V^\xc(k=0)$, together with the kinetic energy, while $V^\xc(k=\pi)$ plays an important role in determining the satellite peaks. One can also write the interaction term as a direct product in momentum domain,
\begin{eqnarray}
&&\sum_{k'}V^\xc(k-k',t)G(k',t)=\nonumber\\
&&V^\xc(0,t)G(k,t)+Y(k,t)G(k,t),
\end{eqnarray}
which gives explicitly the solution for the Green function:
\begin{subequations}\label{eq:Gek_hub}
\begin{align}&&G(k,t>0)=G(k,0^+)e^{-i\varepsilon_kt}e^{-i\int_0^tdt'V^\xc(0,t')}\nonumber\\
&&\times~ e^{-i\int_0^tdt'Y(k,t')},\label{eq:Gek_hub_a}\\
&&G(k,t<0)=G(k,0^-)e^{-i\varepsilon_kt}e^{i\int_t^0dt'V^\xc(0,t')}\nonumber\\
&&\times~ e^{i\int_t^0dt'Y(k,t')}. \label{eq:Ghk_hub_b}
\end{align}
\end{subequations}
One can then use an $N$-site cluster with twisted boundary conditions \citep{tbc1990} to parameterize $G(k,0^{\pm})$ and thus  the generalized Vxc in the momentum domain becomes
\begin{eqnarray}
\label{eq:Z}
Z^\text{el}(k,t):=V^\xc(0,t)+Y(k,t).
\end{eqnarray}
\subsection{Extrapolation from finite clusters}
\begin{figure}[!t]
\centering
\includegraphics[scale=0.6]{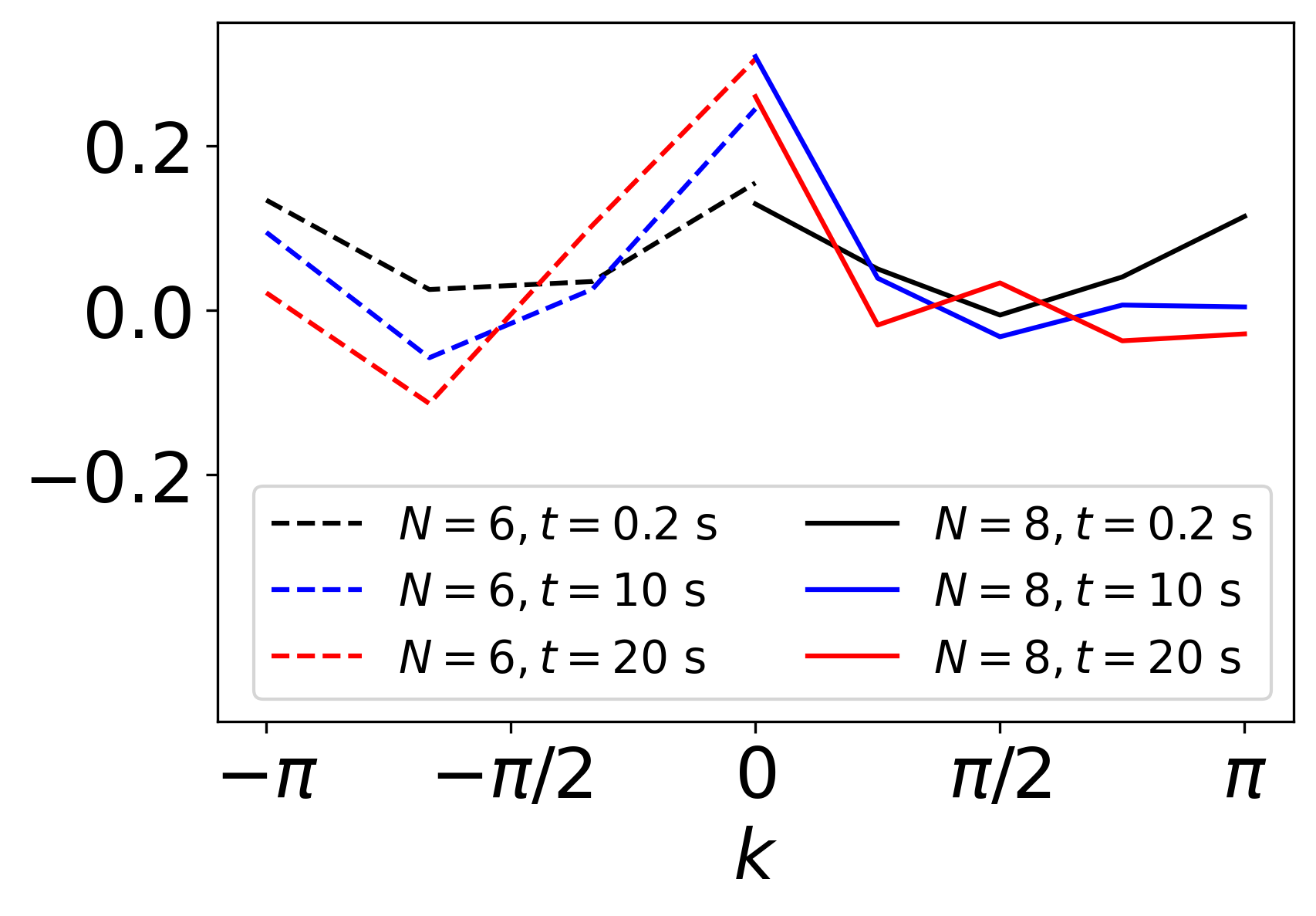}
\includegraphics[scale=0.6]{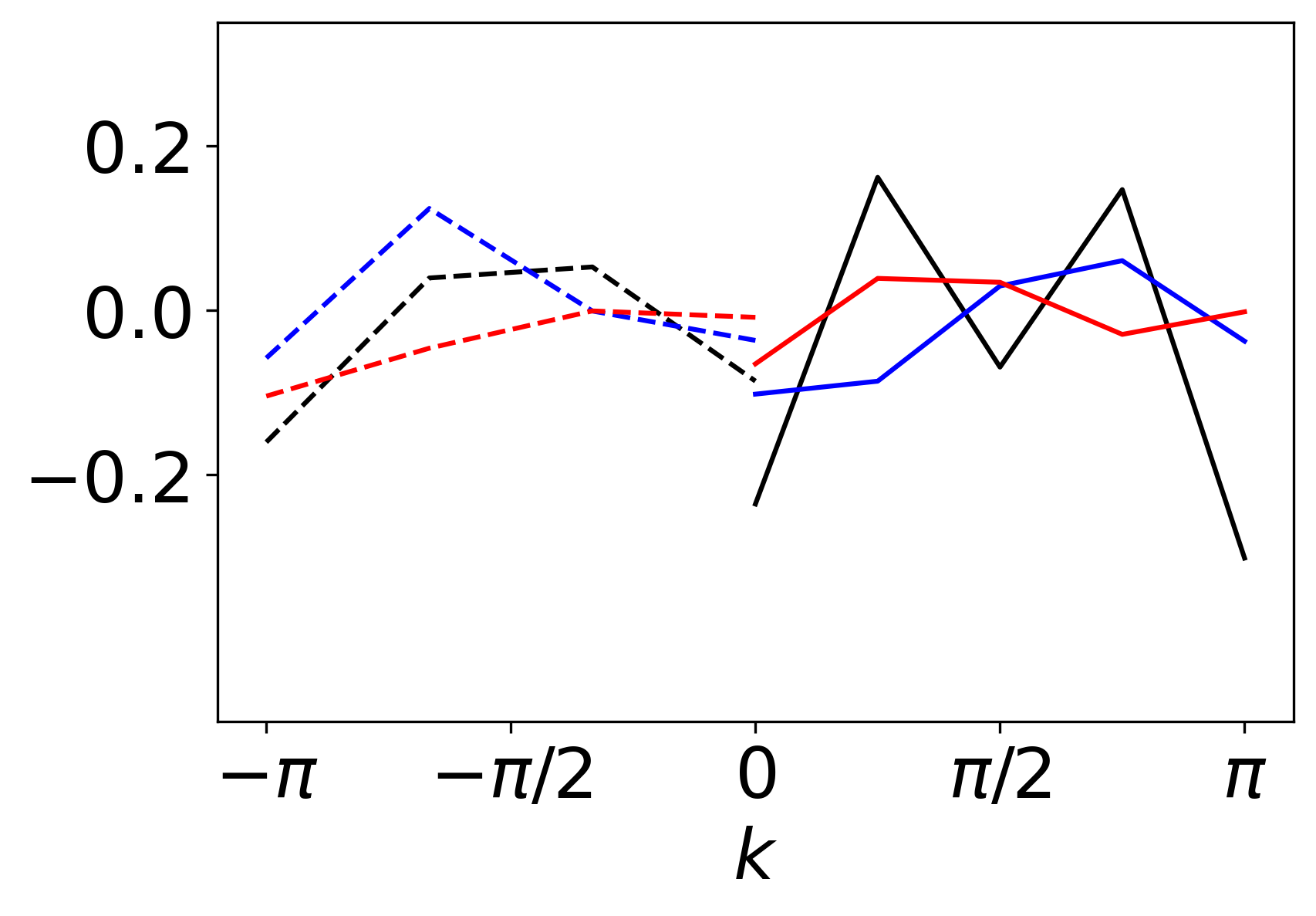}
\caption{$V^\xc(k)$ of finite Hubbard ring, $U=7.74,\Delta=1$, in the unit of $U$. Top panel: real part. Bottom panel: imaginary part. $V(k,t)=V(-k,t)$ and $V(k,t)=-V(-k,-t)$.}
\label{fig:V_hub}
\end{figure}

The Vxc of clusters with 6 and 8 sites and with periodic boundary condition was computed using ED. The Hubbard $U$ was chosen to be $7.74$ with $\Delta=1$, to allow for comparisons with previous work and the DDMRG results from the literature. In contrast to the dimer case, the cluster Vxc exhibits multiple sharp peaks as a function of time $t$. Time snapshots of Vxc as a function of $k$ are shown in Fig.~\ref{fig:V_hub}. For $t\simeq 0$, we have that $V^\xc(k,t)\approx V^\xc(\pi-k,t)$, but such behavior is unseen during the time evolution. The particle-hole symmetry leads to $V^\xc(k,-t)=-V^\xc(k,t)$, and the increase of cluster size from $N=6$ to 8 does not change qualitatively the characteristics of $V^\xc$ as a function of $k$.
\begin{figure*}[t]
\centering
\includegraphics[scale=0.42]{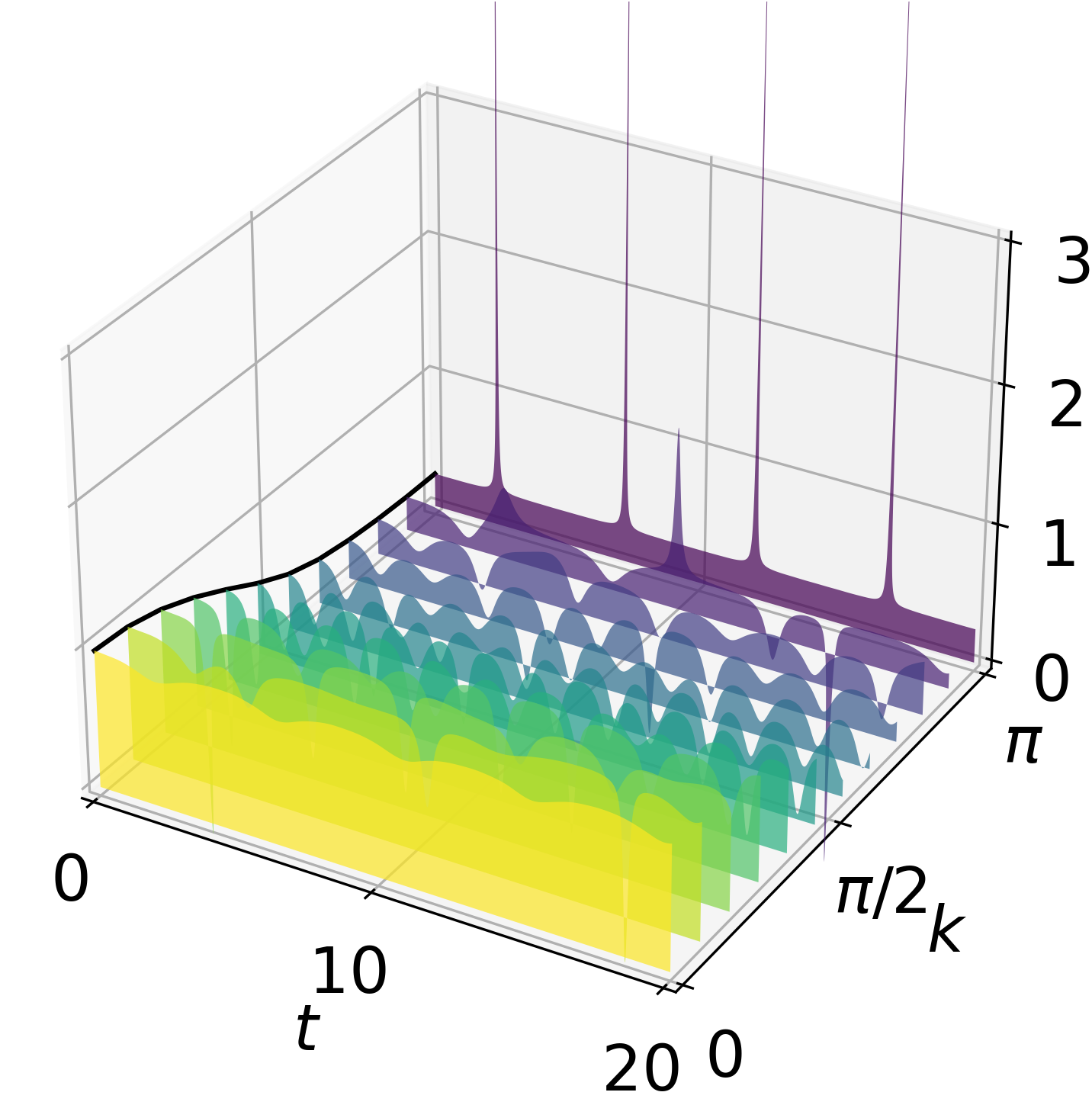}
\includegraphics[scale=0.42]{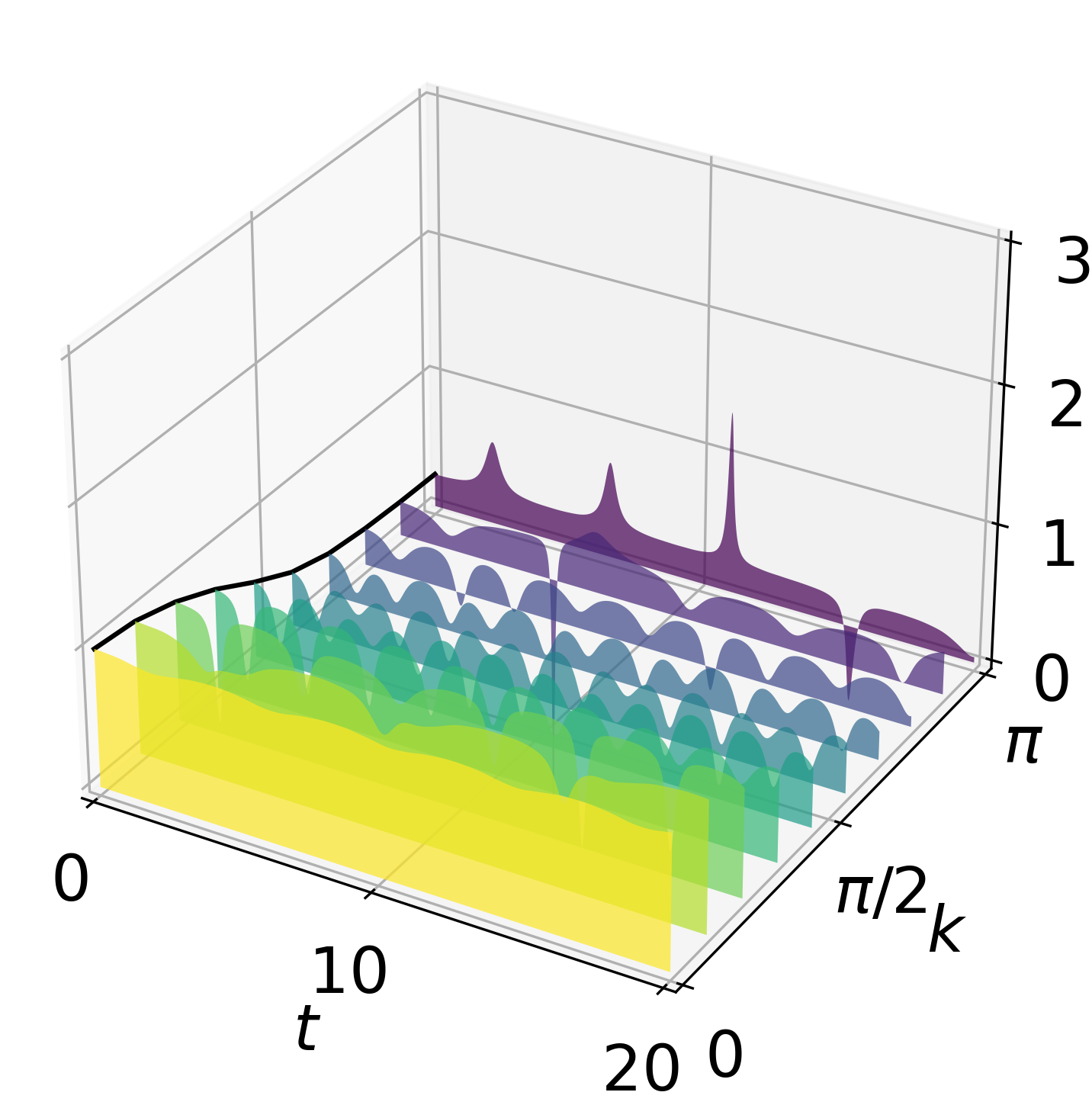}
\includegraphics[scale=0.32]{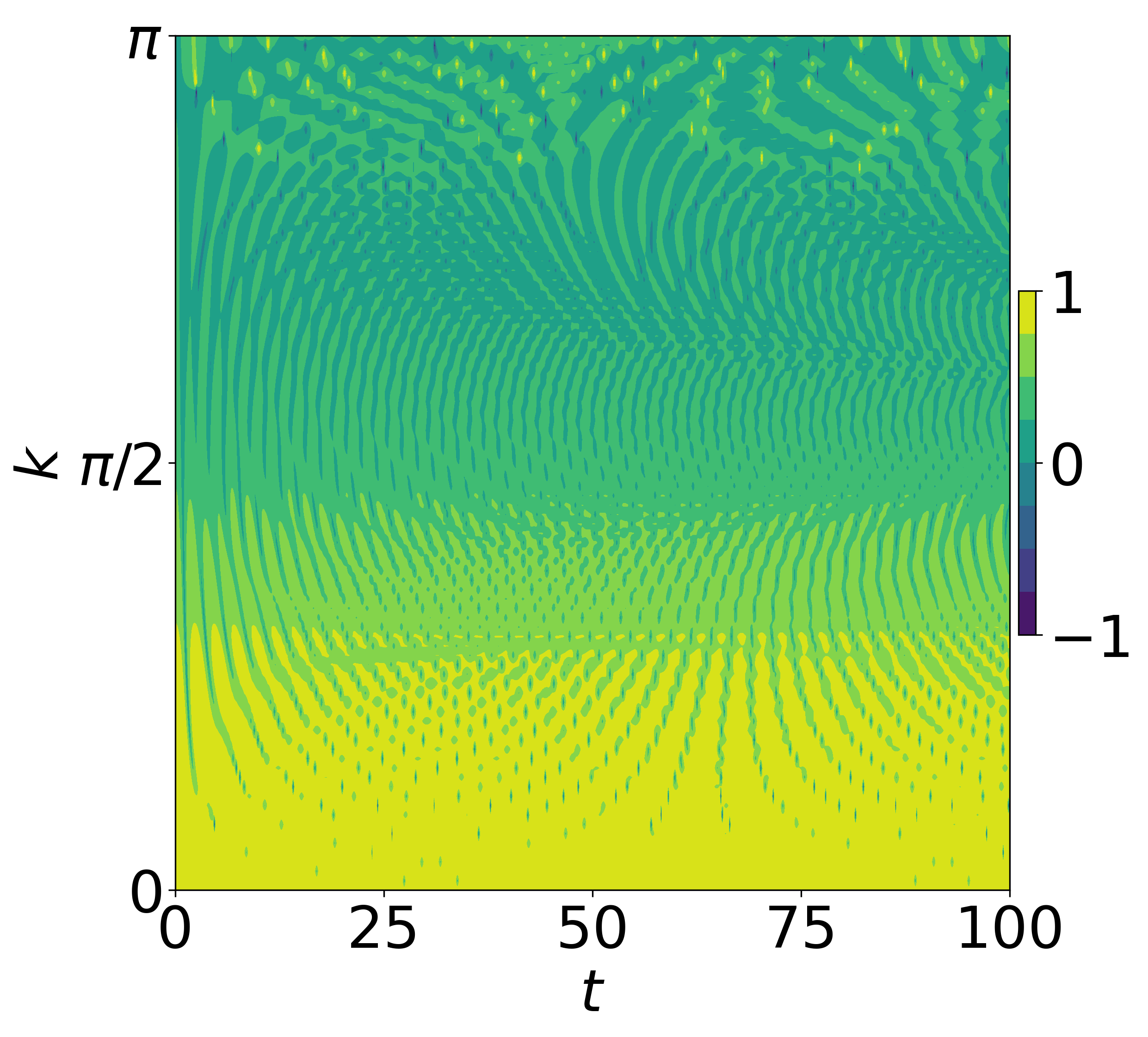}
\caption{Real part of $Z^\text{el}(k,t)$ of finite Hubbard chain with twisted boundary condition, $U=7.74,\Delta=1$, in the unit of $U$. Left and middle panel: with shorter time scale and fewer $k$-points. $N=8,6$, respectively. Right panel: $N=6$, longer time and more $k$-points, peaks out of the color scale not shown. $Z^\text{el}(-k,t)=Z^\text{el}(k,t)$ and $Z^\text{el}(k,-t)=-Z^\text{el}(\pi-k,t)$. For a discussion of the negative peak in the middle panel,
see the main text.}
\label{fig:Z_hub6}
\end{figure*}
The dynamical properties of Vxc can be better illustrated  through $Z^\text{el}$, a generalisation of Vxc in the momentum basis defined in Eq.~\eqref{eq:Z}. Due to degeneracy, $Z^\text{el}(-k,t)=Z^\text{el}(k,t)$ and, because of particle-hole symmetry, $Z^\text{el}(k,-t)=-Z^\text{el}(\pi-k,t)$. To improve the simulation of $Z^\text{el}$, we use a cluster with twisted boundary condition, that 
provides larger $k$-point sampling. The real part of $Z^\text{el}(k,t)$ with twisted boundary condition is shown in Fig.~\ref{fig:Z_hub6}: 
for small $k$, it oscillates weakly in time (with small amplitude and long period). However, where the bandgap opens ($k\rightarrow\frac{\pi}{2}$), the oscillation of Re$Z^\text{el}$ is more evident. For $k\rightarrow\pi$, Re$Z^\text{el}$ exhibits sharp peaks at certain times. The peaks can be both positive and negative: mathematically, this means that some of the zeros of the Green function are located where the  interaction term (Eq.~\eqref{eq:G2_hub}) has nonzero finite (positive or negative) values. These spiky structures cannot be fitted into a weighted sum of several (but finite in number) oscillations, indicating that a model beyond the single-energy quasiparticle picture is necessary.

Provided with the numerically exact Vxc  for $N=6,8$ clusters, we reconsider the approximate scheme 
proposed in the previous work based on the Hubbard dimer ($N=2$) \citep{tor2022}. The
dimer admits two $k$-points ($k=0,\pi)$, with the corresponding approximate values for Vxc given by
\begin{subequations}\label{Vxc_dim1}
    \begin{align}
&V^\xc(k=0,t)\approx\frac{\alpha U}{2},\label{eq:subeq1}\\
&V^\xc(k=\pi,t)\approx\frac{\alpha U}{2}(1-\alpha^2)e^{-i2\Delta t}.\label{eq:subeq2}
     \end{align}
\end{subequations}
Here, the constant $\alpha$ depends only on $\frac{U}{\Delta}$ (the explicit dependence relation is shown in Appendix \ref{app:dimer} together with a
summary of the properties of the Vxc obtained from the Hubbard dimer) and $2\Delta$ in the exponential represents
the main excitation energy. 
In what follows, we use Eqs.~\eqref{eq:subeq1} and \eqref{eq:subeq2} to compute the hole part of the spectral function, with 
the particle part obtainable via the particle-hole symmetry
$A^\text{e}(k,\omega)=A^\text{h}(\pi-k,-\omega)$.
When $|k|\leq\frac{\pi}{2}$, the hole part  of the Green function given by the dimer model is
\begin{eqnarray}
\label{eq:Gh_dimer}
&&G^\text{h}(k,\omega)=\frac{1}{\omega-\omega^\text{h}_k-i\eta}[1-\mathcal{V}^{xc}(\omega)]\\
&&\mathcal{V}^{xc}(\omega)=\frac{1}{N}\sum_{k'}^{\text{occ}}\frac{V^\xc(\pi,0)}{\omega-[\varepsilon_{k'}-V^\xc(0)-2\Delta]-i\eta},
\end{eqnarray}
where $\eta$ is a broadening factor. The spectral function of $G^\text{h}$ has a main peak at $\omega^\text{h}_k$,  determined by $V^\xc(k=0)$ and by the kinetic energy: $\omega^\text{h}_k=\varepsilon_k-V^\xc(0)$. 
The term $\mathcal{V}^{xc}(\omega)$ gives rise to a continuous satellite {region}. Its relative weight to the main peak is $V^\xc(\pi,0)$, and its lower/upper boundaries are given by the minimum/maximum occupied state kinetic energy
\begin{subequations}
\label{eq:wlu}
\begin{align}
\omega^\text{h,lower}_k&=\varepsilon_{0}-V^\xc(0)-2\Delta \label{eq:wl}\\
\omega^\text{h,upper}_k&=\varepsilon_{\frac{\pi}{2}}-V^\xc(0)-2\Delta.\label{eq:wu}
\end{align}
\end{subequations}
The dimer model \citep{tor2022} managed to capture the main structure of the hole spectra of the Hubbard lattice, but can be improved in several aspects: The main peak position given by the model is just the kinetic energy $\varepsilon_k=-2\Delta\cos k$ plus a constant determined by $U$, while the true $k-$dependence of $\omega^\text{h}$ should be more complicated; the upper and lower boundaries of the satellite part given by the model are independent of $k$, which is also an oversimplification.  Rewriting Eq.~\eqref{eq:Ghk_hub_b} in the spirit of a factorization into a main-peak and a satellite term, 
\begin{eqnarray}
G(k,t<0)=G(k,0^-)e^{-i(\varepsilon_k+Z^\text{h,main}_k)t}\times\nonumber\\
e^{i\int_t^0dt'Z^\text{h,sat}(k,t')},
\end{eqnarray}
where $Z^\text{h,main}_k+Z^\text{h,sat}(k,t)=Z^\text{el}(k,t)$ for $t<0$,
one can see that: i) A momentum-dependent static term, $Z^\text{h,main}_k$, which is not present in the dimer model, together with $\varepsilon_k$, determines the main peak; ii) the dispersion of $\omega^\text{h,lower}$ and $\omega^\text{h,upper}$ can be explained by the satellite term $Z^\text{el,sat}(k,t')$. Compared with Fig.~\ref{fig:Z_hub6}, $Z^\text{h,main}_k$ is seen to be the time-independent part around which $Z^\text{el}(k,t)$ oscillates; and $Z^\text{h,sat}(k,t)$ represents a series of excitation energies. The spike-like Re$Z(k,t)$ for $k\rightarrow0,t<0$ is a consequence of multiple excitation energies and large satellite peaks, while the less oscillatory Re$Z(k,t)$ for $k\rightarrow\pi,t<0$ explains the lack of strong satellites of the hole spectral functions $A^\text{h}(k\rightarrow\pi,\omega)$.

\begin{figure}[t]
\centering
\includegraphics[scale=0.35]{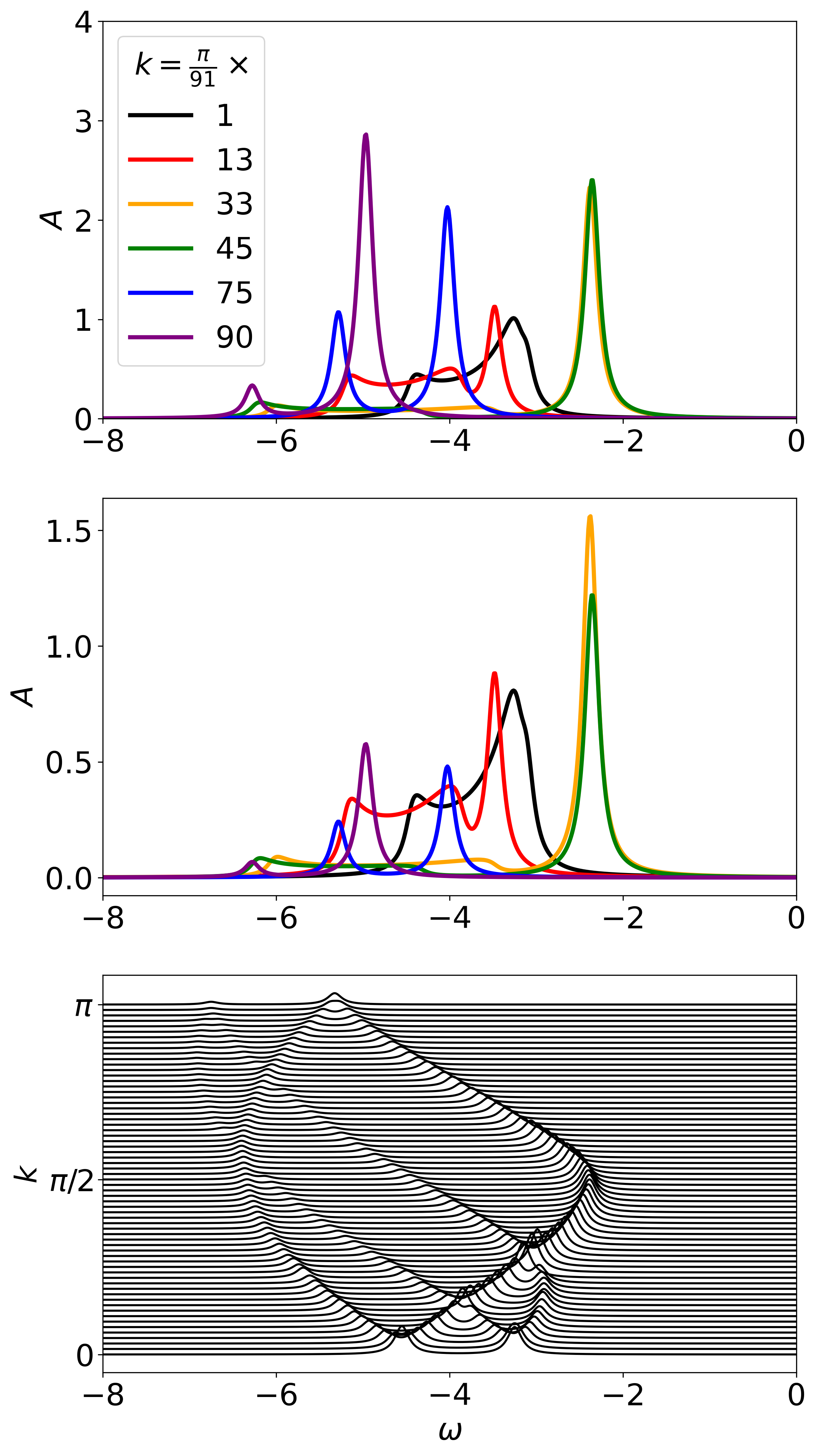}
\caption{Momentum-resolved hole part spectral function $A^h(k,\omega)$ for $U=7.74,\Delta=1$. For $k<\frac{\pi}{2}$, the parameters $l_{1,2,3}$ are determined using the peak locations of the eight-site twisted boundary condition cluster spectrum. For $k>\frac{\pi}{2}$, $Z^\text{el}$ of the eight-site twisted boundary condition cluster is used via Eq.~\eqref{eq:Zel_ansatz} to calculate $A^h$. Top (middle) panel: $k$-points chosen to compare with DDMRG results, without (with) renormalized weight. Bottom panel: the satellite structure is approximated with two peaks at the satellite region boundaries, in order to get clearer dispersion branches. The $k$ values are $\frac{\pi}{64}\times 0,1,2,\cdots,64$. The locations of the spinon branch ($0<k<\pi/2, -3.5<\omega<-2$), the holon branches, and the lower boundary of the holon-spinon continuum ($\pi/2<k<\pi, \omega<-6$) are close to the DDMRG result \citep{tdmrg2007}. For the spinon branch, we have $\omega(k=0)=-3.25$, which differs from the DDMRG result ($\approx -3$) because the finite cluster gives in general larger band gap. In all calculations, we set the
broadening parameter $\eta=0.1$.}
\label{fig:A_from8}
\end{figure}

Taking advantage of the physical picture given by the dimer model, we include the correction to the occupied $k$ values by adding a set of momentum-dependent parameters, $l_{1,2,3}$, such that i) $\alpha\rightarrow\alpha l_1(k)$, ii) the main excitation determining the satellite boundaries (Eq.~\eqref{eq:wl}, \eqref{eq:wu}) becomes $2\Delta\rightarrow2\Delta l_2(k)$, 
and the effective kinetic energy in the summation of Eq.~\eqref{eq:Gh_dimer} becomes $\varepsilon_{k'}\rightarrow -2\Delta\cos k'l_3(k)$.
The parameterized dispersion relations of the key frequencies are
\begin{subequations}
\begin{align}
&\omega^\text{h}_k=-2\Delta \cos k-\frac{\alpha U}{2}l_1(k)\\
&\omega^\text{h,lower}_k=-2\Delta [l_3(k)+l_2(k)]-\frac{\alpha U}{2}l_1(k)\\
&\omega^\text{h,upper}_k=-2\Delta l_2(k)-\frac{\alpha U}{2}l_1(k).
\end{align}
\end{subequations}
Thus, the hole part bandwidth for  a given momentum, the satellite width, and the band gap respectively are 
\begin{subequations}
\begin{align}
&\omega^\text{h}_k-\omega^\text{h,lower}_k=2\Delta \Big[l_2(k)+l_3(k)-\cos k\Big],\\
&\omega^\text{h,upper}_k-\omega^\text{h,lower}_k=2\Delta l_3(k),\\
&E_\text{g}=\alpha Ul_1(\frac{\pi}{2}).
\end{align}
\end{subequations}
This means that the main peak location, the bandwidth, and the satellite region width from cluster calculations can be used to determine the parameters $l_{1,2,3}$, which are then used to calculate the lattice spectral functions $A(k,\omega)$ for $k<\frac{\pi}{2}$. For $k>\frac{\pi}{2}$, where the dimer model gives zero weight for the hole part spectrum, the cluster results show that the corresponding Vxc can be approximated with a single-energy excitation, 
\begin{eqnarray}
\label{eq:Zel_ansatz}
Z^\text{el}(k>\frac{\pi}{2},t<0)\approx \mathcal{A}_ke^{-i\omega^\text{el}_kt}+\mathcal{B}_k,
\end{eqnarray}
where the parameters $\mathcal{A},\mathcal{B}$ and $\omega^\text{el}_k$ are estimated from cluster result, which is similar to the treatment for the spinon Vxc (Eq.~\eqref{eq:sp_quasip}).
Combing the $l_{1,2,3}$-involved occupied region and the $\mathcal{A},\mathcal{B},\omega^\text{el}$-involved unoccupied region, the hole part spectral function can now be calculated for the whole Brillouin zone. The spectral functions for selected $k$ values are shown in Fig.~\ref{fig:A_from8}. 

Compared with the dimer model, the cluster Vxc-based parametrization improves the agreement with DDMRG in several aspects. Specifically,
i) The missing weights for unoccupied $k$ points appear when using as input a cluster Vxc. ii) The main peak positions (and thus the bandgap value as well) are more accurate. In fact, the bandgap value from the dimer model, $\alpha U$, shows a discrepancy with the Bethe ansatz exact value at small $U$, do to the lack of long-range screening effects. Using a cluster Vxc, however, removes the disagreement. iii) Both boundaries and relative weight of the satellite structure are better described by the cluster Vxc and its momentum-basis generalization $Z^\text{el}$; iv) The total weight of the hole/electron part cannot be renormalized within the dimer model, because the non-interacting Green function used in the dimer model can only fix the total spectral weight: $\int d\omega A^\text{h}(k,\omega)=\theta(k_F-k)$. With a cluster Vxc, using $\ex{\hat{c}^\dagger_k\hat{c}_k}$, we can rescale the total spectral weight for each $k$ value.

Yet, the main peak $\omega^\text{h}$:s in Fig.~\ref{fig:A_from8} is in general lower than the one from DDMRG. This can be understood  as
due to the band gap narrowing upon increasing the numbers of site (the eight-site cluster we used leads to the overestimation of the gap and thus 
of the main peak position).

We conclude our discussion of the Hubbard chain, by considering its spectral functions in real space that we obtain starting from
those in the momentum domain:
\begin{eqnarray}
A(r,\omega)=\frac{1}{2\pi}\int dk A(k,\omega)e^{ikr}
\end{eqnarray}
where $r=0,1,2,\cdots$ is in units  of lattice parameter. $A(r,\omega)$ describes the correlation strength between two space points separated by $r$, at a given energy $\omega$. The local case $A(r=0,\omega)$ corresponds to the density of states. 

\begin{figure}[t]
\centering
\includegraphics[scale=0.45]{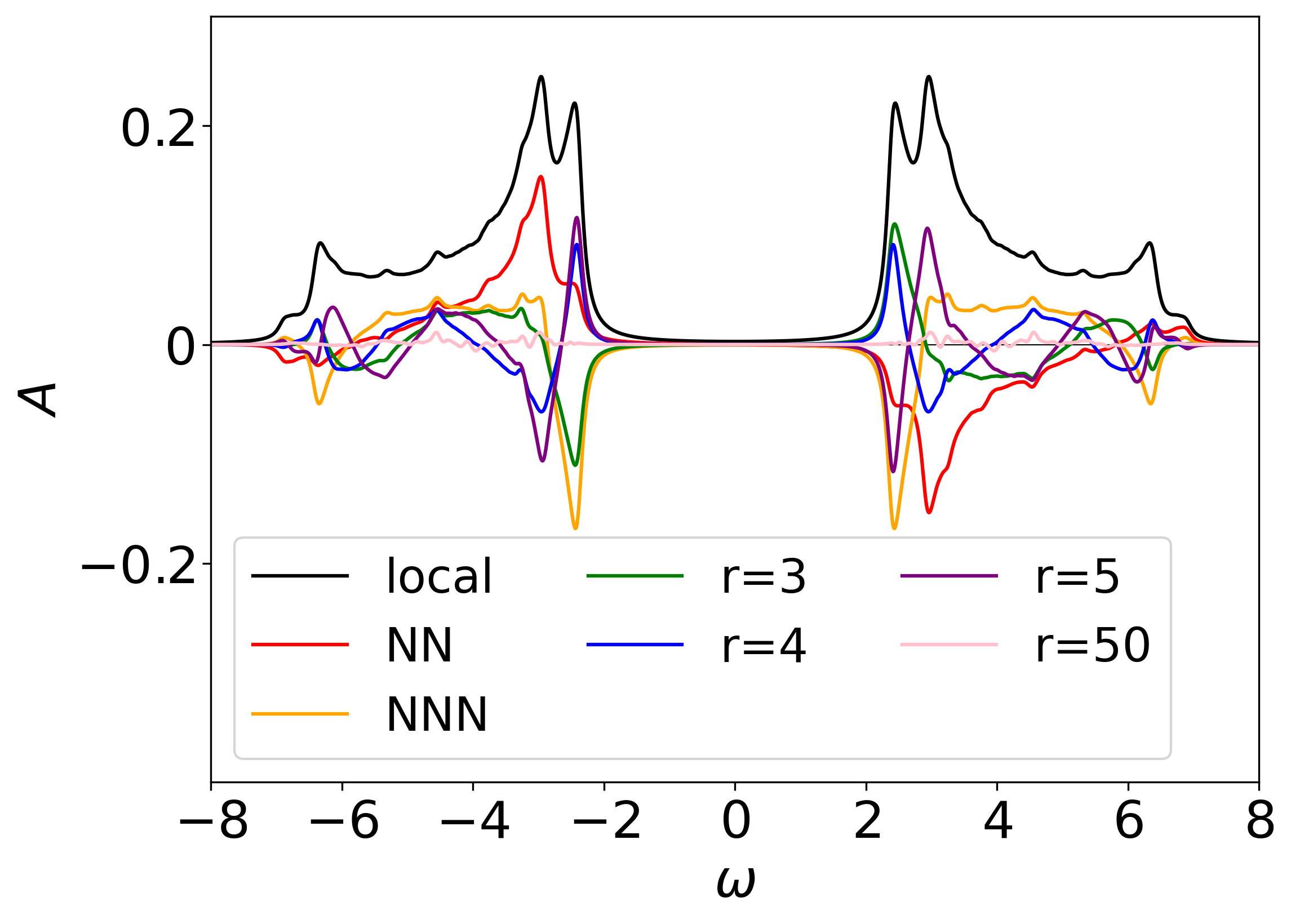}
\caption{Spatial spectral functions, calculated with the eight-site Vxc as kernel (for the six-site case, see the Appendix) for $U=7.74,\Delta=1$ and broadening $\eta=0.1$. 64 $k$-points are used to approximate the $k$-integral, according to the Chadi-Cohen method \citep{chadi1973}.}
\label{fig:Ar_hub}
\end{figure}

Results for $A(r,\omega)$ with a eight-site kernel are shown in Fig.~\ref{fig:Ar_hub}, whilst those from  a six-site kernel with different $U$ and $r$ are reported in the appendix.
The cluster Vxc result for $A(r=0,\omega)$ shows better agreement with DDMRG than the dimer model. 
Also, the NN spectral weight at positive energy is  predominantly negative, and for $r\geq 2$, $A(r,\omega)$ exhibits nodal structures. Concerning the role of electronic correlations, spatial spectral functions with different $U$ values become qualitatively alike 
at large repulsion ($U>4$), but the band gap value keeps increasing with $U$. Finally, spectral functions calculated with eight-site and six-site kernels 
are qualitatively similar (see the appendix for the six-site case), with similarities in the overall shape and in the number of nodes. However, the estimated value of the band gap improves on increasing the cluster size.


\section{Vxc from Hubbard and Heisenberg models: a comparative discussion}\label{sec:compare}
It is well known that the 1D spin$-\frac{1}{2}$ AFM Heisenberg model becomes equivalent to the 1D half-filled Hubbard
model in the large $U$ regime  \citep{fazekas1999,nolting2009}. After having discussed Vxc in the two models separately, it can be useful
to look at both models together using as perspective the behavior of Vxc in such limit.
Meanwhile, $Z^\text{el}$ and $Z^\text{sp}$ do not show a direct asymptotic behavior $Z^\text{el}\big|_{U\rightarrow\infty}=Z^\text{sp}$, because they are coupled to the single-particle Green function (Eq.~\eqref{eq:GF_hub_def}) and single spin-flipping Green function (Eq.~\eqref{eq:GF_heis_def}) respectively. For the Hubbard model, the term corresponding to $Z^\text{sp}$ is coupled to the two-particle Green function $\ex{\mathcal{T}\big[\hat{c}_{p\uparrow}^\dagger(t)\hat{c}_{p\downarrow}(t)\big]\big[\hat{c}_{q\downarrow}^\dagger\hat{c}_{q\uparrow}\big]}$. Equation of motion of higher order Green function needs to be solved for the Hubbard model to calculate the `higher order Vxc' that is comparable with the spinon Vxc under large repulsion. This means the Vxc formalism for Heisenberg model, having a similar sum rule (Eq.~\eqref{eq:sum_rule}), reduces the difficulty in deriving the equation of motion and improves the interpretability via the quasiparticle picture. 

\begin{figure}[t]
\centering
\includegraphics[scale=0.4]{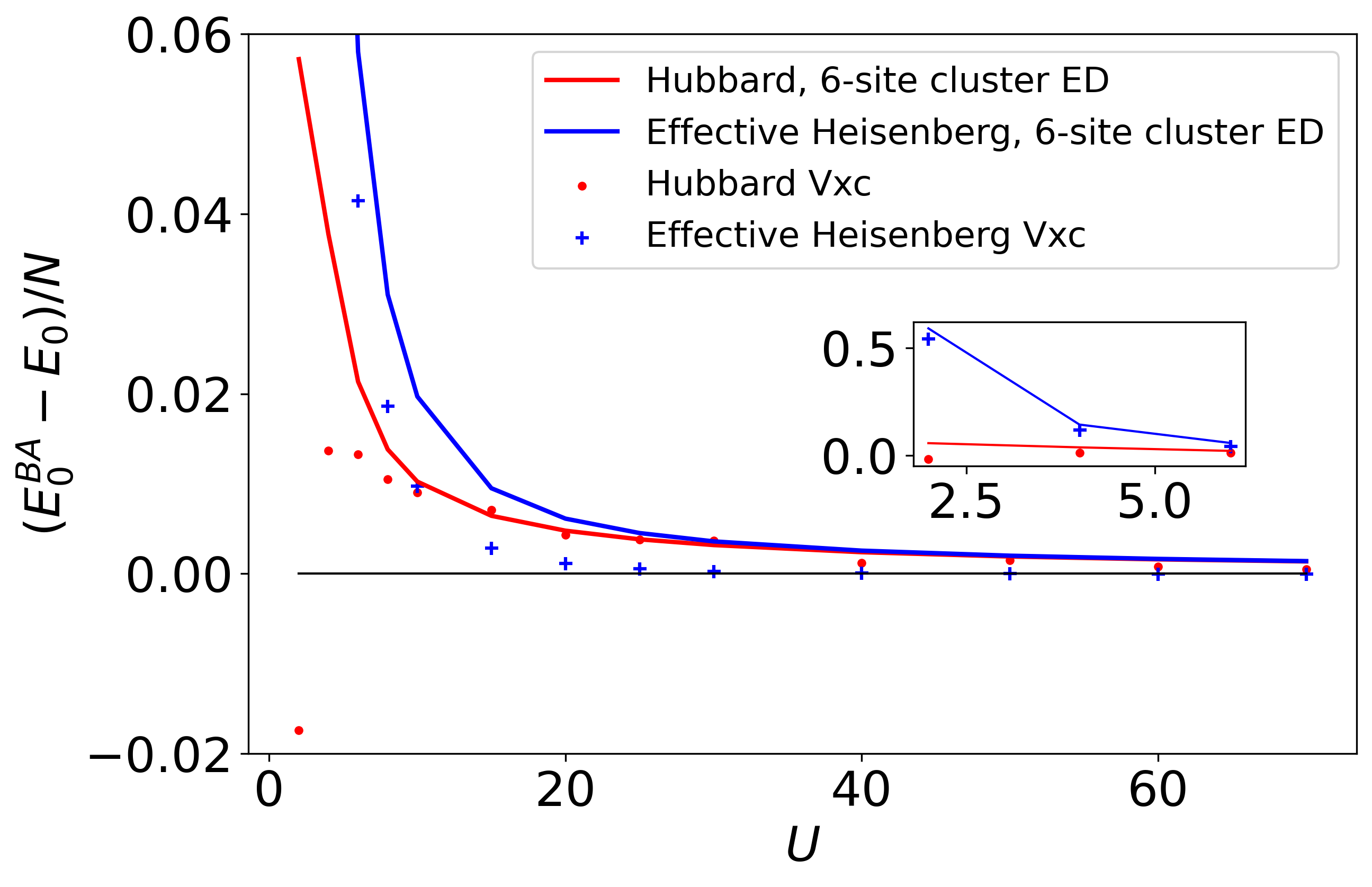}
\caption{Difference between the exact Bethe ansatz $\frac{E_0^\text{Hub}}{N}$ and the Vxc-based results for i) the 1D Hubbard model  and ii) the 1D AFM Heisenberg model, and  the ED results for iii) a six-site Hubbard cluster and iv) a six-site Heisenberg cluster. For both models, Vxc  is extrapolated from a six-site kernel.}
\label{fig:E0_comp}
\end{figure}

Instead of solving the higher order Green function, we consider a more modest task of comparing the lattice ground state energies for the two models. In the large $U$ limit \citep{nolting2009},
\begin{eqnarray}
\label{eq:largeU}
\lim_{U\rightarrow\infty}\frac{E_0^\text{Hub}}{N}=\frac{1}{U}(4\frac{E_0^\text{Heis}}{N}-1)
\end{eqnarray}
where $E_0^\text{Hub}$ is the ground state energy of a $N$-site Hubbard ring with $\Delta=1$, and $E_0^\text{Heis}$ is the ground state energy of a $N$-site AFM Heisenberg ring with $J=-1$. Both energies can be calculated from the Green function via
\begin{eqnarray}
\frac{E_0^\text{Heis}}{N}=\frac{3}{2}\ex{S_1^+(t=0^+)S_2^-},
\end{eqnarray}
and
\begin{eqnarray}
\frac{E_0^\text{Hub}}{N}=-\big[2\ex{\hat{c}_{1\uparrow}^\dagger\hat{c}_{2\uparrow}(t=0^-)}-i\partial_t\ex{\hat{c}_{1\uparrow}^\dagger\hat{c}_{1\uparrow}(0^-)}\big].
\end{eqnarray}
In the frequency domain,
\begin{eqnarray}
\frac{E_0^\text{Heis}}{N}&=&\frac{3i}{4\pi}\int G^\text{sp}(r=1,\omega)d\omega,\\
\frac{E_0^\text{Hub}}{N}&=&\frac{i}{2\pi}\int \Big[2G^\text{el}(r=1,\omega)-\omega G^\text{el}(r=0,\omega)\Big]d\omega.\nonumber\\
\end{eqnarray}
To perform a comparison, we compute the ground state energy of the Hubbard lattice in two ways: i)  by directly using the electron Vxc at different $U$ values, and ii) by calculating $E_0^\text{Heis}$ for a $J=-1$ Heisenberg lattice with the spinon Vxc, to be then used
in the effective $E_0^\text{Hub}$ of Eq.~\eqref{eq:largeU}. The differences between the results from these two prescriptions and the exact Bethe ansatz solution are shown in Fig.~\ref{fig:E0_comp}. The $E_0$ results from ED for a six-site ring are also shown as a reference. 

For $U<10$, the repulsion strength is not large enough for Eq.~\eqref{eq:largeU} to be valid, leading to a discrepancy between the total energies for the two lattice models. However, in such region, $E_0^\text{Hub,Vxc}$ (red dots) is already close to the exact Bethe ansatz value, and the difference gets smaller on increasing $U$. For $U>30$, the ED results for the two models converge, meaning that the large repulsion limit is reached. The Vxc-based energies $E_0$ for the two models also converge to the exact Bethe ansatz value. 

However, the effective Vxc-based Heisenberg result is rather accurate, with absolute error less than $10^{-4}$: this can be understood as a result of i) using the two-spinon upper and lower boundaries in the extrapolation, and ii) adjusting the $\mathcal{B}$ parameter from the cluster within the zero spin gap picture. In contrast, the Vxc-based Hubbard result is extrapolated without a good reference, and is more affected by the finite size effects. Thus, the difference with the Bethe ansatz result is larger. 

As an overall remark,
the comparative analysis of this Section shows the versatility of the Vxc approach across different lattice models, with results that are consistent with trends and benchmarks from other methods.

\section{Conclusion and outlook}
\label{sec:con}
We have presented a novel exchange correlation potential (Vxc) formalism for
the one-dimensional antiferromagnetic Heisenberg model, and derived a general new sum rule for spin systems. 
Our spin-formulation is a tailored extension of a previously introduced general framework for many-body systems that include
both charge and spin degrees of freedom. Together with the new formulation, we have also devised a procedure
to obtain, from a Vxc extracted from small finite clusters, an extrapolation to the thermodynamical limit.
This procedure to access Vxc, originally devised for spin systems, has also permitted us to revisit and improve the treatment of 
the half-filled one-dimensional Hubbard model, a system already considered in earlier work within the Vxc approach. 
For both the 1D AFM Heisenberg model and the 1D Hubbard model, the static exchange term of Vxc was derived and shown to exhibit model distinctive properties. 
For the 1D AFM Heisenberg model, the static exchange term corresponds to the lower boundary of the two-spinon spectrum. For the Hubbard model, the local $U$ leads to a constant $V^\text{x}$, which cancels the Hartree potential. 

For both models, the spectral functions calculated within the Vxc approach show favourable agreement with DDMRG and with experimental results. Furthermore, a single-energy quasiparticle picture can be used to explain the dynamics of the spinon Vxc for the 1D AFM Heisenberg model and the unoccupied/occupied part of the hole/electron Vxc for the 1D Hubbard model. Finally, we showed how the Vxc formalism captures the equivalence of the two models in the large $U$ limit, by a comparative analysis via the lattice ground state energies. 

In conclusion, our results indicate that the Vxc formalism provides an alternative way of calculating single-particle Green function which is (computationally) cost-beneficial but also physically well defined. Looking forward, we plan to apply such dimensionality- and interaction-insensitive scheme to models of increasing complication and higher dimensionality. At the same time, we intend to explore ways to devise Vxc approximations with the goal of 
improving over the local-(spin)-density approximation of density functional theory, 
in a progression toward a first principle implementation for real materials.

\section{Acknowledgments}
F.A. gratefully acknowledges
financial support from the Knut and Alice Wallenberg  
Foundation (KAW 2017.0061) and the Swedish Research Council 
(Vetenskapsrådet, VR 2021-04498\_3). C.V. 
gratefully acknowledges financial support from the Swedish Research Council 
(VR 2017-03945 and  2022-04486).

\appendix
\begin{figure*}
\centering
\includegraphics[scale=0.4]{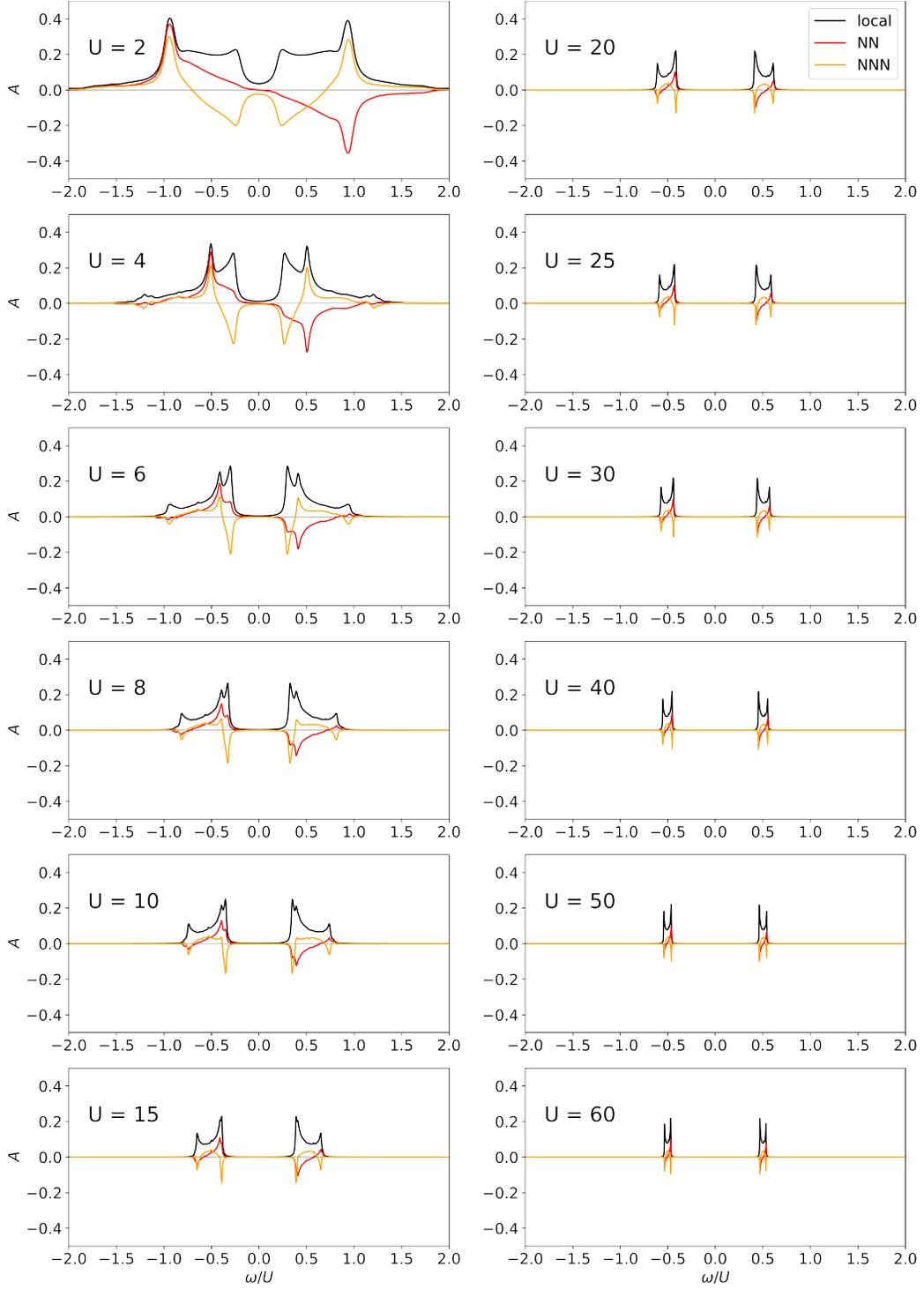}
\caption{Spatial spectral function of Hubbard chain, calculated with six-site kernel.}
\label{fig:ArU6}
\end{figure*}

\section{Sum rule and exchange term of Heisenberg chain}
\label{app:sum_rule}
Equation of motion of the Heisenberg model is
\begin{eqnarray}
i\partial_t G_{pq}(t)+iF_{pq}(t)=2\delta_{pq}\delta(t)\ex{S^z_p}
\end{eqnarray}
where the interaction term is
\begin{eqnarray}
F_{pq}(t)=-J\sum_{\delta}[\ex{p,p+\delta;q}-\ex{p+\delta,p;q}],
\end{eqnarray}
and 
\begin{eqnarray}
\ex{l,p;q}&:=&\ex{\mathcal{T}\hat{S}_l^z(t^+)\hat{S}_p^+(t)\hat{S}_q^-(0)}.
\end{eqnarray}
The correlator $g_{lpq}(t)$ and the exchange-correlation hole are defined to fulfill:
\begin{eqnarray}
\ex{l,p;q}&=&iG_{pq}(t)g_{lpq}(t)\ex{\hat{S}_l^z}\\
\rho^\xc_{lpq}(t)iG_{pq}(t)&=&-\ex{l,p;q}+\ex{\hat{S}_l^z}iG_{pq}(t)\\
\rho^\xc_{lpq}(t)&=&-\big[g_{lpq}(t)-1\big]\ex{\hat{S}_l^z}.
\end{eqnarray}
For $t>0$,
\begin{eqnarray}
\label{eq:G2sp_pt}
\sum_l\ex{l,p;q}=S^ziG_{pq}(t),
\end{eqnarray}
and for $t<0$,
\begin{eqnarray}
\label{eq:G2sp_nt}
\sum_l\ex{l,p;q}&=&\sum_l\Big[\ex{\hat{S}_q^-(0)\hat{S}_p^+(t)\hat{S}_l^z(t)}+\ex{\hat{S}_q^-(0)\hat{S}_p^+(t)}\delta_{pl}\Big]\nonumber\\
&=&(1+S^z)iG_{pq}(t).
\end{eqnarray}
Eq. \eqref{eq:G2sp_pt} and \eqref{eq:G2sp_nt} can be written in a compact form as
\begin{eqnarray}
\sum_l\ex{l,p;q}=\big[\theta(-t)+S^z\big]iG_{pq}(t).
\end{eqnarray}
Therefore the correlator fulfills
\begin{eqnarray}
\sum_liG_{pq}(t)\big[g_{lpq}(t)-1\big]\ex{\hat{S}_l^z}&=&\sum_l\ex{l,p;q}-\sum_l\ex{\hat{S}_l^z}\nonumber\\
&=&\theta(-t)iG_{pq}(t),
\end{eqnarray}
from which the sum rule can be retrieved:
\begin{eqnarray}
\sum_l\rho_{lpq}^\xc(t)=-\theta(-t).
\end{eqnarray}

The exchange term of spinon Vxc can be derived from the
variational method
\begin{equation}
\dd{G_{pq}(t)}{\varphi_l(t^+)}=-\ex{l,p;q}+\ex{\hat{S}^z_l}iG_{pq}(t).
\end{equation}
so the interaction term can be written as
\begin{eqnarray}
F_{pq}(t)=-J\sum_{\delta}\Big[\dd{G_{pq}(t)}{\varphi_{p+\delta}(t^+)}-\dd{G_{p+\delta,q}(t)}{\varphi_{p}(t^+)}\nonumber\\
+\ex{S^z_p}iG_{p+\delta,q}(t)-\ex{S^z_{p+\delta}}iG_{pq}(t)\Big]
\end{eqnarray}
According to the definition of Vxc,
\begin{eqnarray}
V^\xc_{pp,qq}(t)iG_{pq}(t)=-J\sum_{\delta}\Big[\dd{G_{pq}(t)}{\varphi_{p+\delta}(t^+)}-\dd{G_{p+\delta,q}(t)}{\varphi_{p}(t^+)}\Big]\nonumber\\
\end{eqnarray}
and 
\begin{eqnarray}
V^\text{H}_p&=&J\sum_\delta\ex{S^z_{p+\delta}},\\
V^\text{F}_p&=&-J\ex{S^z_{p}},
\end{eqnarray}
the equation of motion can be rewritten as
\begin{eqnarray}
[i\partial_t-V^\text{H}_p]G_{pq}(t)
-\sum_\delta V^\text{F}_pG_{p+\delta,q}(t)\nonumber\\
-V^\xc_{pp,qq}(t)G_{pq}(t)=2\delta_{pq}\delta(t)\ex{S^z_p}
\end{eqnarray}
Considering
\begin{eqnarray}
\dd{G(1,2)}{\varphi(3)}=-\int d4d5G(1,4)\dd{G^{-1}(4,5)}{\varphi(3)}G(5,2)
\end{eqnarray}
and the lowest order of the vertex function is
\begin{eqnarray}
\dd{G^{-1}(4,5)}{\varphi(3)}=-\delta(4-5)\delta(4-3),
\end{eqnarray}
one gets the exchange part
\begin{eqnarray}
&V^\text{x}_{pp,qq}(t)iG_{pq}(t)
=-J\times\nonumber\\
&\sum_{\delta=\pm 1}\Big[G_{p,p+\delta}(0^-)G_{p+\delta,q}(t)-G_{p+\delta,p}(0^-)G_{pq}(t)\Big].\nonumber\\
\end{eqnarray}
\section{Analytic Vxc of four-site Heisenberg chain}
\label{app:spVxc}
To compute the Green function for positive time, 
\begin{eqnarray}
G_{pq}(t)&=\ex{\Psi|e^{iHt}\hat{S}_p^+e^{-iHt}\hat{S}_q^-|\Psi}
\end{eqnarray}
where $|\Psi\rangle$ is the ground state, 
one needs to use a complete set of eigenstates $|n\rangle$ which give nonzero weight elements $\ex{n|\hat{S}_q^-|\Psi}$. For an  even number of sites and AFM coupling, the total z-spin of $|\Psi\rangle$ is zero, which means that the states $\{n\rangle\}$ are in the $S^z=-1$ sector. Labeling the eigenenergy corresponding to state $|n\rangle$ with $E^{-}_n$, the Green function can be written as
\begin{eqnarray}
\!\!\!\!\!\!\!\!G_{pq}(t>0)&=\sum_n e^{-i(E_n^--E^0)t}\ex{\Psi|\hat{S}_p^+|n\rangle\langle n|\hat{S}_q^-|\Psi},
\end{eqnarray}
and the high order term for positive time is
\begin{eqnarray}
\!\!\!\!\!\!\!\!\!\!\!\! \ex{l,p;q}_{t>0}=\sum_n e^{-i(E_n^--E^0)t}\ex{\Psi|\hat{S}_l^z\hat{S}_p^+|n\rangle\langle n|\hat{S}_q^-|\Psi}.
\end{eqnarray}
By diagonalizing the Hamiltonian in the $S^z=0$ and $S^z=-1$ sectors, one respectively gets $\lbrace|\Psi\rangle; E^0\rbrace$ and $\lbrace|n\rangle;E_n^-\rbrace$, and thus the weight elements $\ex{n|\hat{S}_q^-|\Psi}$ and $\ex{n|\hat{S}_p^-\hat{S}_l^z|\Psi}$. Out of the 4 states of $|n\rangle$, only 3 of them give nonzero $\ex{n|\hat{S}_q^-|\Psi}$. Explicitly, the time factors are
\begin{eqnarray}
f_1&=&e^{-i(E_0^--E^0)t}=e^{iJ(\frac{\sqrt{3}-\sqrt{2}+1}{2})t},\\ 
f_2&=&e^{-i(E_1^--E^0)t}=e^{iJ(\frac{\sqrt{3}+1}{2})t}\\
f_3&=&e^{-i(E_2^--E^0)t}=e^{iJ(\frac{\sqrt{3}+\sqrt{2}+1}{2})t}.
\end{eqnarray}
The independent elements of $V^\xc$ in orbital basis can be calculated with $V^\xc_{pq}(t)=\frac{F_{pq}(t)}{iG_{pq}(t)}$:
\begin{widetext}
\begin{eqnarray}
V^{\xc}_{11}&=&-J\frac{(\frac{(xy+x)(xy+x+2y)}{a_+^2})f_1+(x^2+x)f_2+(\frac{(xy-3x)(xy-3x+2y-4)}{a_-^2})f_3}{(\frac{xy+x+2y}{a_+})^2f_1+x^2f_2+(\frac{xy-3x+2y-4}{a_-})^2f_3}\\
V^{\xc}_{22}&=&-J\frac{(\frac{2(x+1)(xy+x+2)}{a_+^2})f_1+(x^2+x)f_2+(\frac{2(x+1)(xy-3x-2)}{a_-^2})f_3}{(\frac{xy+x+2}{a_+})^2f_1+x^2f_2+(\frac{xy-3x-2}{a_-})^2f_3}\\
V^{\xc}_{12}&=&-J\frac{(-\frac{(xy+x)(xy+x+2)}{a_+^2})f_1-(x^2+x)f_2-(\frac{(xy-3x)(xy-3x-2)}{a_-^2})f_3}{(-\frac{(xy+x+2y)(xy+x+2)}{a_+^2})f_1-x^2f_2-(\frac{(xy-3x+2y-4)(xy-3x-2)}{a_-^2})f_3}\\
V^{\xc}_{13}&=&-J\frac{(\frac{(xy+x)(xy+x+2)}{a_+^2})f_1-(x^2+x)f_2+(\frac{(xy-3x)(xy-3x-2)}{a_-^2})f_3}{(\frac{(xy+x+2y)(xy+x+2)}{a_+^2})f_1-x^2f_2+(\frac{(xy-3x+2y-4)(xy-3x-2)}{a_-^2})f_3}\\
V^{\xc}_{14}&=&-J\frac{-(\frac{(xy+x)(xy+x+2y)}{a_+^2})f_1+(x^2+x)f_2-(\frac{(xy-3x)(xy-3x+2y-4)}{a_-^2})f_3}{-(\frac{xy+x+2y}{a_+})^2f_1+x^2f_2-(\frac{xy-3x+2y-4}{a_-})^2f_3}\\
V^{\xc}_{23}&=&-J\frac{-(\frac{2(x+1)(xy+x+2)}{a_+^2})f_1+(x^2+x)f_2-(\frac{2(x+1)(xy-3x-2)}{a_-^2})f_3}{-(\frac{xy+x+2}{a_+})^2f_1+x^2f_2-(\frac{xy-3x-2}{a_-})^2f_3}.
\end{eqnarray}
\end{widetext}
where the constant factors $x,y$ and $a_{\pm}$ are defined in the main text.
The terms in `bonding-antibonding' basis are then
\begin{widetext}
\begin{eqnarray}
V_{BB,BB}^{\xc}=\frac{1}{16}\Big[&&2(V_{11}^{\xc}+V_{14}^{\xc}+V_{22}^{\xc}+V_{23}^{\xc})
+4(V_{12}^{\xc}+V_{13}^{\xc})\Big]\\
V_{BC,CB}^{\xc}=\frac{1}{16}\Big[&&2(V_{11}^{\xc}-V_{14}^{\xc}+V_{22}^{\xc}-V_{23}^{\xc})
+4(V_{12}^{\xc}-V_{13}^{\xc})\Big]\\
V_{BA,AB}^{\xc}=\frac{1}{16}\Big[&&2(V_{11}^{\xc}-V_{14}^{\xc}+V_{22}^{\xc}-V_{23}^{\xc})
-4(V_{12}^{\xc}-V_{13}^{\xc})\Big]\\
V_{BD,DB}^{\xc}=\frac{1}{16}\Big[&&2(V_{11}^{\xc}+V_{14}^{\xc}+V_{22}^{\xc}+V_{23}^{\xc})
-4(V_{12}^{\xc}+V_{13}^{\xc})\Big]
\end{eqnarray}
\end{widetext}

\section{The dependence of the $\alpha$ parameter on $\frac{U}{\Delta}$ in the Hubbard dimer model}
\label{app:dimer}
The equations in this subsection are rewritten from the Hubbard dimer work \citep{tor2022}. With a two-site open ends chain, the half-filled Hubbard Hamiltonian Eq.\eqref{eq:def_hub} can be analytically solved, given the analytic bonding ($k=0$) and anti-bonding ($k=\pi$) Vxc:
\begin{subequations}\label{eq:Vxc_dim0}
\begin{align}
&V^\xc(k=0,t>0)=\frac{\alpha U}{2}\frac{1-\alpha^2 e^{-i4\Delta t}}{1-\alpha^4 e^{-i4\Delta t}}\\
&V^\xc(k=\pi, t>0)=\frac{\alpha U}{2}\frac{(1-\alpha^2) e^{-i2\Delta t}}{1-\alpha^4 e^{-i4\Delta t}},
\end{align}
\end{subequations}
where
$\alpha=\frac{1-\kappa}{1+\kappa}$, and $\kappa=\frac{1}{4}\Big(\sqrt{(\frac{U}{\Delta})^2+16}-\frac{U}{\Delta}\Big)$.
After neglecting the higher excitation term $e^{-i4\Delta t}$ in Eq.~\eqref{eq:Vxc_dim0}, the approximated dimer Vxc in the main text (Eq.~\eqref{Vxc_dim1}) is obtained.

\bibliography{bib}

\end{document}